\documentclass[a4paper,11pt]{article}
\usepackage{graphicx,amssymb,amstext,amsmath,mathabx}
\usepackage{color, yfonts, tcolorbox}
\usepackage{floatflt}
\usepackage{tikz,pgfplots}
\usepackage{float}
\pgfplotsset{compat=1.3}
\usetikzlibrary{spy}
\usepackage{hyperref}
\hypersetup{
    colorlinks=true,
    linkcolor= blue,
    filecolor=magenta,      
    urlcolor=blue,
}

\definecolor{cbl}{rgb}{0,0,1}                

\usepackage{mathtools}

\newcommand{\bc}{\begin{center}}
\newcommand{\ec}{\end{center}}
\def\ba#1{\begin{array}{#1}\displaystyle}
\newcommand{\ea}{\end{array}}

\newcommand{\beq}{\begin{equation}}
\newcommand{\eeq}{\end{equation}}
\newcommand{\beqa}{\begin{eqnarray}}
\newcommand{\eeqa}{\end{eqnarray}}

\newcommand{\bi}{\begin{itemize}}
\newcommand{\ei}{\end{itemize}}

\newcommand{\p}{\partial}

\begin{document}
\begin{titlepage}
\vspace{0.2cm}
\begin{center}

{\large{\bf{Generalised Hydrodynamics of Particle Creation and Decay}}}

\vspace{0.8cm} 
{\small \text{Olalla A. Castro-Alvaredo${}^{\heartsuit}$, Cecilia De Fazio${}^{\diamondsuit}$, Benjamin Doyon${}^{\spadesuit}$, and Aleksandra A. Zi\'o\l{}kowska ${}^{\clubsuit} $}}

\vspace{0.8cm}
{\small ${}^{\heartsuit\,\diamondsuit}$} \small Department of Mathematics, City, University of London, 10 Northampton Square EC1V 0HB, UK\\
\vspace{0.2cm}
\small
${}^{\spadesuit}$  Department of Mathematics, King's College London, Strand WC2R 2LS, UK \\
\vspace{0.2cm}

\small
\hspace{-0.2cm}${}^{\clubsuit}$   Rudolf Peierls Centre for Theoretical Physics, University of Oxford, Parks Road, Oxford OX1 3PU, UK

\end{center}
\medskip
\medskip
\medskip
\medskip
Unstable particles rarely feature in conjunction with integrability in 1+1D quantum field theory. However, the family of homogenous sine-Gordon models provides a rare example where both stable and unstable bound states are present in the spectrum whilst the scattering matrix is diagonal and solves the usual bootstrap equations. In the standard scattering picture, unstable particles result from complex poles of the $S$-matrix located in the unphysical sheet of rapidity space. Since  they are not part of the asymptotic spectrum, their presence is only felt through the effect they have on physical quantities associated either to the theory as a whole (i.e.~scaling functions, correlation functions) or to the stable particles themselves (i.e.~energy/particle density). In two recent publications, the effect of unstable particles in different out-of-equilibrium settings has been studied. It has been shown that their presence is associated with specific signatures in many quantities of physical interest. A good way to select those quantities is to adopt the generalised hydrodynamic approach and to consider the effective velocities and particle densities of the stable particles in the theory. For an initial state given by a spacial gaussian profile of temperatures peaked at the origin, time evolution gives rise to particle and spectral particle densities that exhibit hallmarks of the creation and decay of unstable particles. While these signatures have been observed numerically elsewhere, this paper explores their quantitative and qualitative dependence  on the parameters of the problem. We also consider other initial states characterised by ``inverted gaussian" and ``double gaussian" temperature profiles.

\vspace{1cm}

\medskip

\vfill

\noindent 
${}^{\heartsuit}$ o.castro-alvaredo@city.ac.uk\\
${}^{\diamondsuit}$ cecilia.de-fazio.2@city.ac.uk\\
${}^{\spadesuit}$ benjamin.doyon@kcl.ac.uk\\
${}^{\clubsuit}$ aleksandra.ziolkowska@sjc.ox.ac.uk\\

\hfill \today

\end{titlepage}

\section{Introduction}
Consider a two-body scattering process in a 1+1D  quantum field theory described by a diagonal scattering matrix $S_{ab}(\theta)$ where $a, b$ label the particle species, $\theta$ is their rapidity difference and ``diagonal" means that the in-coming and out-going particles are identical. If this scattering matrix has a simple pole of the form $\theta=i \pi u$ with $0<u<1$ then this can be interpreted as the formation of a stable bound state $c$ of mass $m^2_c=m^2_a+ m_b^2+2m_a m_b \cos \pi u$. This stable bound state will then be part of the asymptotic spectrum of the theory, on exactly the same footing as particles $a, b$.  In contrast, if the $S$-matrix has a pole of the form $\theta=\sigma-i \bar{\sigma}$ with $0<\bar{\sigma}<\pi$ and $\sigma \in \mathbb{R}^{\geq 0}$ this is associated with the formation of an unstable bound state or unstable particle with mass $M$ and decay width $\gamma$ given by the Breit-Wigner formula \cite{Breit-Wigner}:
\beq
2M^2=\sqrt{A^2+B^2}+A \qquad \mathrm{and}\qquad  \frac{\gamma^2}{2}=\sqrt{A^2+B^2}-A\,,
\label{BW1}
\eeq
for 
\beq
A=m_a^2+m_b^2+2m_a m_b \cosh\sigma \cos\bar{\sigma}\qquad \mathrm{and} \qquad  B=2m_a m_b \sinh\sigma \sin\bar{\sigma}\,.
\label{BW2}
\eeq
It is well-known that the structure of the scattering matrices and correlation functions of integrable quantum field theory (IQFT) is severely constrained both by standard physical requirements (i.e.~unitarity and crossing symmetry) and by the presence of an infinite number of conserved quantities $Q_i$ with $i=0, \pm 1, \pm 2 \ldots$ However, this does not forbid the formation of both stable and unstable bound states. Nonetheless, while stable bound states are a common feature of most interacting IQFTs, unstable bound states are a rare occurrence. 

The single largest family of theories containing both kinds of bound states are the homogeneous sine-Gordon (HSG) models. They can be seen as massive perturbations of a critical Wess-Zumino-Novikov-Witten model \cite{WZNW1,WZNW2,WZNW3,WZNW4,WZNW5} associated to cosets $G_k/U(1)^{r_g}$ where $G$ is some simply-laced algebra, $k$ is the level (an integer), and $r_g$ is the rank of $G$. Despite their name, they have little in common with the sine-Gordon model. In particular, their scattering matrices are diagonal. 

 The HSG-models were first named and studied by the ``Santiago de Compostela" group in a series of papers in the late 90s where their classical and quantum integrability were established \cite{hsg,ntft}, their particle spectrum determined  \cite{FernandezPousa:1997iu}, and an exact scattering matrix eventually proposed \cite{smatrix}. The scattering matrix was then tested extensively by employing the thermodynamic Bethe ansatz (TBA) \cite{tba1,tba2}  and the form factor approach \cite{KW,SmirnovBook}. The TBA of these models was studied in detail in \cite{ourtba,CastroAlvaredo:2002nv,Dorey:2004qc}, while the form factors of the local operators of some theories were constructed in \cite{CastroAlvaredo:2000em,CastroAlvaredo:2000nk}. The effect of the presence of unstable particles in the renormalisation group flow of several quantities was also explored using form factor techniques in \cite{CastroAlvaredo:2000ag,CastroAlvaredo:2000nr} and the mass-coupling relation was determined in \cite{Bajnok1,Bajnok2}.

This paper is part of an ongoing study \cite{ourUP,tails} of the out-of-equilibrium dynamics of these theories and, in particular, of the precise signatures the presence of unstable particles leaves in dynamical quantities of physical interest. The main tool we employ in order to compute such quantities is the generalised hydrodynamic approach (GHD), a leading method in the study of the dynamics of many-body quantum systems, particularly integrable ones \cite{ourhydro,theirhydro,Brev}. GHD builds upon the much greater understanding of such systems that has been gained over the past decade  \cite{20,Eisert,NERev2,NERev3,NERev4,NERev5,NERev6,NERev7,Brev,CEM} and also on the special role that integrable models have come to play within these developments. Indeed, the results of the quantum Newton's cradle experiment \cite{kinoshita}  showed that the intersection of dimensionality and integrability gives rise to a distinct kind of dynamics, one in which there is no long-term thermalization. This result was later related to the presence of infinitely many conserved quantities in integrable models (as mentioned earlier)  and the fact that all these quantities may be involved in determining the long time dynamics, giving rise to the concept of generalised Gibbs ensembles (GGEs) \cite{Rigol}. As a consequence, quantum integrable models do not thermalise (to Gibbs ensembles) but instead relax to GGEs. 

Both here and in two previous publications \cite{ourUP, tails} we focus our study on one specific theory within the HSG family.  This is in fact the simplest non-trivial example and is known as the $SU(3)_2$-homogeneous sine-Gordon model (associated with the coset $SU(3)_2/U(1)^2$, where two stable and one unstable particle are present.  Labelling the stable particles by an index $\pm$, the two-body scattering matrix of the theory has the simple form,
\beq
S_{\pm \pm}(\theta)=-1 \qquad \mathrm{and} \qquad S_{\pm \mp}(\theta)=\pm \tanh\frac{1}{2}\left(\theta\pm \sigma-\frac{i\pi}{2}\right)\,,
\eeq
and the associated differential scattering phases (sometimes referred to as kernels in the TBA context) are
\beq
\varphi_{\pm\pm}(\theta)=0 \qquad \mathrm{and}\qquad \varphi_{\pm\mp}(\theta)=\mathrm{sech}(\theta\pm\sigma)\,.
\label{sphases}
\eeq
$S$-matrix parity is broken as $S_{+-}(\theta) \neq S_{-+}(\theta)$ and $\varphi_{+-}(\theta) \neq \varphi_{-+}(\theta)$ for $\sigma\neq 0$. The kernel still has spatial parity invariance if we associate the particles' charges with their spatial parity, $\varphi_{+-}(\theta) = \varphi_{-+}(-\theta)$, and this will be used to simplify the discussion below.
S-matrix parity breaking means that interaction between particles is maximised when their rapidity difference is $\theta=\mp \sigma$ and the kernel reaches its maximum value $+1$. As a consequence, many functions resemble free fermion solutions for rapidities that are far from the values $\mp\sigma$ whereas they show features of interaction for rapidities near these values. This is the reason why we speak about a ``free fermion peak" of the spectral densities in later sections.

The $S$-matrix has a pole at $\theta=\sigma-\frac{i\pi}{2}$ and reduces to a system of two free fermions for $\sigma\rightarrow +\infty$. From the RG viewpoint, the theory describes the massless flow between a theory of two free fermions ($c=1$) and a non-trivial interacting conformal field theory of central charge $c=6/5$ \cite{ourtba}.

From the Breit-Wigner formulae (\ref{BW1})-(\ref{BW2}) it follows that the associated unstable particle has mass $M$ such that for $\sigma\gg 1$ and given that the masses of particles $\pm$ are equal $m_{\pm}:=m$, the mass of the unstable particle and the decay width $\gamma$ both scale as
\beq
M\sim \gamma \sim m e^{\frac{\sigma}{2}}\,.
\label{tres}
\eeq 
This energy scale plays the role of a threshold for the formation of the unstable particle, in effect separating the free fermion from the interacting regime.

Parity breaking and the presence of the unstable particle have many interesting effects both at and away from equilibrium. These effects are very clear when investigated in the context of GHD, that is, when looking at functions such as the effective velocities of stable particles, their associated densities and spectral densities. Definitions of those functions can be found in the next section (see Eqs. (\ref{qjpm})-(\ref{eff})). At thermal equilibrium, the following main features were found \cite{ourUP}:
\begin{itemize}
\item  {\bf Unstable Particle Threshold:} Particle densities and effective velocities develop non-trivial patterns as the temperature changes, which can be explained by the formation of unstable particles at energies beyond the threshold (\ref{tres}). Furthermore, beyond the same threshold and due to parity breaking, there are non-zero energy currents for the individual particles $\pm$. These are equal and of opposite sign ensuring that the total current is zero, as expected at equilibrium.
\item  {\bf Monotonicity Changes:} For temperatures $T$ beyond the threshold (\ref{tres}) the spectral densities of the stable particles develop three local maxima (instead of the usual two) centered around rapidities $\pm \log(2T), \mp \log(2T)$ and $\mp\sigma\pm\log(2T)$ for particles $\pm$. We have named these peaks the free fermion, interacting and subsidiary peak. 

\item {\bf Unstable Particle Creation:}  The subsidiary peak of particle ($+$) and part of the interacting peak of particle ($-$) can be interpreted as representing densities of mutually interacting particles, that is, in effect, a constant density of unstable particles. These interacting particles are also co-moving (in rapidity/phase space), that is, they have the same effective velocities. 
\end{itemize}
Away from equilibrium, some of the features above generalise in a natural fashion while new dynamics also comes into play, notably the dynamics of unstable particle decay. In \cite{ourUP} we considered the partitioning protocol, where two copies of the system, thermalised at different temperatures are connected at time zero and allowed to evolve for a long time. In this context the main features of the equilibrium physics follow through albeit now depending on two distinct temperatures. For instance currents associated to either particle are still non-zero but the $\pm$ contributions no longer add up to zero and net ballistic currents emerge. Similarly, a finite density of unstable particles can be read off from the structure of the spectral densities and effective velocities, but the latter are no longer simply related by a parity transformation.
\medskip

More relevant for the present paper are the results of \cite{tails} where we considered instead an initial condition characterised by a gaussian temperature profile of the form 
\beq
T(x)=T_a+(T_m-T_a)e^{-x^2} \qquad \mathrm{with} \qquad \sigma>T_m>m e^{\frac{\sigma}{2}}>T_a\geq 0\,.
\label{gaussian}
\eeq
where $x$ is the space coordinate. 
In this case we found:
\begin{itemize}
\item {\bf Unstable Particle Decay:} As at equilibrium, the spectral particle densities have three local maxima which are (roughly) the time-evolved versions of those found at equilibrium at temperature $T_m$. Under time evolution in the absence of a bath ($T_a=0$) the original subsidiary peak decays, leaving behind a tail of slower particles. A tail is also observed in the particle densities, while the shape of the effective velocities is altered for values of $\theta$ around the location of the subsidiary peak.

    \item {\bf ``Magnetic Fluid"-Like Effect:} For $T_a\neq 0$ however, the subsidiary peak does not entirely decay. Instead, it partly decays for early times and partly persists for large times. The persistence is directly linked to the presence of the bath and can be interpreted as a novel ``magnetic fluid"-like effect:  the persisting peak is driven over the bath as the bath interacts with the interacting peak of the other particle type. Consequently, even though particles of type $(+)$ inside their persisting peak have a range of effective velocities smaller than $+1$, the peak itself propagates at speed  $+1$, propelled by the interacting peak of particle $(-)$. Nonetheless, this does not represent the full range of behaviours seen in true magnetic fluids as, for instance, particles in the bath at the sort of temperatures that we consider here, do not interact with each other.
\end{itemize} 
The main objective of this paper is to explain in more quantitative and qualitative detail the two phenomena above. We will do this by employing extensive numerics based on the iFluid package \cite{iFluid,iFluid2} already employed in \cite{tails} and some analytical results based on standard TBA and GHD techniques. Our aim is to understand how both the decay (in the absence of a bath) and the persistence (in the presence of a bath) of a residual density of particles of each species are related to the choice of parameters in the problem, that is $T_m, T_a$ and $\sigma$ in (\ref{gaussian}). Regarding the decay process, we focus on three specific quantities:
\bi
\item the maximum height of the subsidiary peak of the spectral density functions $\rho_p^\pm(x,t,\theta)$,
\item the maximum height of the decaying peak of the particle density, after subtraction of the free fermion contribution, $H_0^\pm(t)$. This is in effect the $\theta$-integrated subsidiary peak of the spectral density, and 
\item the $x$-integrated particle density once the free fermion has been subtracted, $Q_0^\pm(t)$.
\ei 
For all quantities, we find that, there are three characteristic time scales. For early times the dynamics is fully determined by the original gaussian profile of temperatures  and this can be easily corroborated both analytically and numerically.
For intermediate times we observe exponential decay. The decay rate is different for each of the functions considered, but in all cases provides a measure of the rate at which the initial density of unstable particles is reduced through their decay as time evolves. For the maximum of the spectral density, the decay rate is explicitly related to the slope of the effective velocity, whereas for the other functions it is a function of the universal ratio $T_m/M$, where $M$ is the mass of the unstable particle. Finally, for late times, the decay either accelerates (in the absence of a bath) or stops (in its presence). Except for early times where some analytical predictions can be made, our analysis is based on numerically solving the GHD equations employing iFluid \cite{iFluid,iFluid2}. 

\medskip

This paper is organised as follows: In Section~\ref{tbaeq} we review the main equations and functions of interest in the context of GHD. In Section~\ref{qdecay} we introduce three functions of particular interest in the context of quantifying the decay/persistence of an effective unstable particle density under time evolution. We perform a numerical study of these three functions and conclude that in all cases there is a substantial time interval for which their decay is exponential. In Section~\ref{magnet} we investigate in detail the properties of the persisting peak of the spectral density that is formed in the presence of a bath, focusing on how its height and volume relate to the temperatures $T_m$ and $T_a$. In Section~\ref{otherthings} we discuss the main features of the spectral density and effective velocities for two new initial conditions, characterised by an inverted gaussian and double gaussian temperature profile. We conclude in Section~\ref{conclu}. In Appendix~\ref{App:A} we discuss some properties of the particle density in the free fermion case. In Appendix~\ref{App:B} we find a universal formula in integrable QFT for the particle density at thermal equilibrium in the limit of high temperature.
Tables of numerical values of the fitting parameters for the many fits presented in the paper are gathered in Appendix~\ref{fitt}. 

\section{Dynamics: Thermodynamics and Hydrodynamics}
\label{tbaeq}
A natural question is: given an initial state characterised by the temperature profile (\ref{gaussian}), how do we compute dynamical quantities for generic values of $x$ and $t$? The quantities we have in mind here are the GGE averages of charge densities $q_i(x,t)$ of conserved charges $Q_i = \int dx q_i(x,t)$, and their currents $j_i(x,t)$ satisfying $\p_t q_i(x,t) + \p_x j_i(x,t)=0$. In this section we review a number of analytical results which are relevant in the study of this problem, following \cite{ourhydro}. As mentioned in the introduction, integrable models do not necessarily thermalise for large times but they locally equilibrate to GGEs.  In such a state, the thermodynamic/dynamic properties of the theory are described by the (generalised) TBA equations \cite{tba1,tba2,FM,Mossel}, which for this model are simply 
\beq
\varepsilon^\pm(x,t,\theta)=w(\theta,x,t)-\varphi_{\pm\mp}\star L^\mp(x,t,\theta)\,,
\label{pseudoen}
\eeq 
with the kernel $\varphi_{\pm\mp}(\theta)$ defined in (\ref{sphases}).
Here $\star$ represents the convolution 
\beq
a \star b(\theta):= \frac{1}{2\pi} \int_{-\infty}^\infty a(\theta-\theta') b(\theta') d\theta'\,,
\label{convo}
\eeq
$L^\pm(x,t,\theta)=\ln(1+e^{-\varepsilon^\pm(x,t,\theta)})$ and the driving term $w(x,t,\theta)$ contains a sum over the one-particle eigenvalues of the conserved charges involved in the GGE. We call these eigenvalues $h_i(\theta)$ with $i=0,1,2,\ldots$ and in this case they are identical for both particle types so they carry no particle indices
\beq
w(x,t,\theta)=\sum_i \beta_i(x,t) h_i(\theta)\,,
\eeq
where $\beta_i(x,t)$ are generalised inverse temperatures, which are fluid-cell dependent. For simplicity, we will take the mass of the particles $m=1$ for the remainder of the paper.  As we can see, all quantities will generally be functions of space and time.

A special property of the equations (\ref{pseudoen}) for our model when considered at equilibrium in a Gibbs ensemble at temperature $T=1/\beta$
\beq
\varepsilon^\pm(\theta)=\beta \cosh\theta-\varphi_{\pm\mp}\star L^\mp(\theta)\,,
\label{pseudoeneq}
\eeq 
is that they can be mapped into 
\beq
\phi^{\pm}(\theta)=\beta \cosh(\theta\mp\frac{\sigma}{2})-(\varphi \star K^{\mp})(\theta)\quad {\rm with} \quad  \varphi(\theta):={\rm sech}\theta\,.
\eeq
for the shifted functions
\beq
\phi^\pm(\theta):=\varepsilon^\pm(\theta\mp \frac{\sigma}{2})\quad \mathrm{and} \quad K^\pm(\theta)=\log(1+e^{-\phi^\pm(\theta)})\,.
\label{shifted}
\eeq
Note that under this shift and a change of variables in the convolution integral, the $\sigma$-dependence of the TBA kernel is eliminated so that it is now only explicit in the TBA driving term.

Then, if $\sigma$ or $\theta$ are large and positive we can approximate
\beq
\phi_{\pm}(\theta)= e^{\mp \theta -\kappa}-(\varphi \star K_{\mp})(\theta)
\eeq
where 
\beq
\kappa=\log(2T)-\frac{\sigma}{2}\,.
\label{kappa}
\eeq 
This shows that for $\sigma$ large enough, the TBA equations and their solutions naturally depend on the variable $\kappa$. This observation extends to the out-of-equilibrium situation, as long as the initial state is defined by a temperature profile, as we shall see later. 
\medskip

From these objects, averages of  the local densities $q_i(x,t)$ are fully fixed by giving the one-particle eigenvalues of the associated conserved charge, $h_i(\theta)$. The averages are obtained by using the ``dressed" quantities $h_i^{\mathrm{dr},\pm}(x,t, \theta)$, which solve the linear integral equations
\beqa
h_i^{\mathrm{dr},\pm}(x,t, \theta)= h_i(\theta)+ \varphi_{\pm \mp} \star g^\mp_i(x,t,\theta)\,,
\label{dress}
\eeqa 
where
\beq
g_i^\pm(x,t,\theta)= h_i^{\mathrm{dr},\pm}(x,t,\theta) n^\pm(x,t,\theta) \qquad \mathrm{and} \qquad n^\pm(x,t,\theta)=\frac{1}{1+e^{\varepsilon^\pm(x,t,\theta)}}\,,
\label{ndefini}
\eeq 
is the occupation function associated to particle $\pm$. In this paper we will focus on just a small number of quantities. The first one is the particle density 
 $\texttt{q}_0(x,t)=\langle q_0 \rangle_{\underline{\beta}}$ which for our model can be expressed as
\beqa 
\texttt{q}_0(x,t)=\sum_{b=\pm} \int_{-\infty}^\infty \frac{d \theta}{2\pi} e^{\rm{dr},b}(x,t,\theta) n^b(x,t,\theta)\,,
\label{qj}
\eeqa 
(here $\underline{\beta}$ is the set of generalised inverse temperatures in the GGE) and we have implicitly used the one-particle eigenvalue $h_0(\theta)=1$ (particle number).
Above, the dressed energy is computed through (\ref{dress}) from its one-particle eigenvalue  $h_1(\theta)=e(\theta)= \cosh \theta$. The formula above allows us to naturally
attribute a density contribution to each particle type. We will define those as
\beqa 
\texttt{q}_0^{\pm}(x,t)= \int_{-\infty}^\infty \frac{d \theta}{2\pi} e^{\rm{dr},\pm}(x,t,\theta) n^\pm(x,t,\theta)\,,
\label{qjpm}
\eeqa 
Two intermediate functions in these expressions are of particular interest, as they possess a clear physical meaning: these are the spectral density (the integrand of (\ref{qjpm})), and the effective velocities, which first appeared in \cite{BEL14},
\beq
\rho_p^\pm(x,t, \theta)=\frac{1}{2\pi}e^{\rm{dr},\pm}(x,t,\theta) n^\pm(x,t,\theta)\,\qquad\mbox{and}\quad
v^{\textrm{eff},\pm}(x,t, \theta)=\frac{p^{\rm{dr},\pm}(x,t,\theta) }{e^{\rm{dr},\pm}(x,t,\theta) }\,,
\label{eff}
\eeq
respectively. Because of the equations satisfied by particles $\pm$ and, specially the symmetry of the kernel $\varphi_{+-}(\theta)=\varphi_{-+}(-\theta)$, we have that the functions above are not fully independent but rather satisfy
\beq
\texttt{q}_0^{+}(x,t)=\texttt{q}_0^{-}(-x,t),\,\, \,\rho_p^+(x,t, \theta)=\rho^-_p(-x,t, -\theta)\,\,\, \,\mathrm{and}\,\, \,\,v^{\textrm{eff},+}(x,t, \theta)=-v^{\textrm{eff},-}(-x,t, -\theta)\,.
\label{parity}
\eeq
Because of these relations, we will focus most of our analysis on particle ($+$), keeping in mind that similar conclusions hold for particle ($-$).

The effective velocity depends on the dressed energy and momentum which has one-particle eigenvalue $h_2(\theta)=p(\theta)=\sinh\theta$.  The spectral density is a conserved quantity, and the spectral density times the effective velocity, is its current. In other words, it satisfies the conservation equation
\beq
\partial_t \rho_p^\pm(x,t,\theta)+\partial_x(v^{\textrm{eff},\pm}(x,t, \theta) \rho_p^\pm(x,t,\theta))=0\,.
\label{cons}
\eeq
In summary, given the functions $n^\pm(x,t,\theta)$ for all values of $x,t$ and $\theta$ we can obtain any of the other functions defined here. The question is then: given the functions $n^\pm(x,0,\theta)$, which in our case are obtained by solving the TBA equations for a thermal state with temperature (\ref{gaussian}), how can we obtain the function $n^\pm(x,t,\theta)$ at later times? 

\medskip
This question can be addressed numerically in a very effective manner by using for instance the iFluid package \cite{iFluid, iFluid2} which has demonstrated its efficacy in a number of examples, including the computation of correlation functions. Indeed, we have used this package to numerically solve our problem in \cite{tails} and will also employ it here for all our numerical results. Details on the modifications of the package that have been needed for this particular model, as well as numerical consistency checks can be found in  Appendix A of \cite{tails}. Note that
 iFluid is a very general package, designed to consider all manner of out-of-equilibrium situations, which means that it is not necessarily the most algorithmically efficient way of numerically solving our problem. For inhomogeneous quenches, this question was first discussed in \cite{DSY} where a solution based on  the method of characteristics was proposed.

\section{Quantifying Decay}
\label{qdecay}
 Recall the natural scale (\ref{kappa}) and the threshold (\ref{tres}) for interaction introduced earlier. Combining both, we can say that the scale $\kappa$ provides a measure of where along the RG-flow the model is. In the out-of-equilibrium situation we need to be more specific though, as we now have two different temperature scales, $T_m$ and $T_a$. Except for subsection \ref{inverted} where the roles of $T_m$ and $T_a$ are reversed, we will generally choose the temperature of the bath to be fixed and low, below the interaction threshold, so that the largest temperature having the strongest influence in the dynamics will be $T_m$. Therefore, when we refer to $\kappa$ here and in later sections we will be talking about the scale:
 \beq
 \kappa=\log(2T_m)-\frac{\sigma}{2}\,.
 \eeq 
 A value $\kappa=0$  corresponds exactly to the midpoint between the low energy behaviour, described by two free fermions, and the UV behaviour, characterised by the presence of the unstable particle. The larger $\kappa$ is, the closer we are to the UV fixed point. 
In summary we can say that there are three interesting regimes
\beq
\begin{array}{cl}
    \kappa>0 & \rm{Interacting\, Regime} \\
     \kappa=0  & \rm{Interaction \, Threshold}\\
     \kappa<0 & \rm{Non-Interacting\, Regime}\\
\end{array}
\eeq
In our paper \cite{tails} we chose parameters $2T_m=e^7$ and $\sigma=10$ such that $\kappa=2$. This is above the threshold for the formation of unstable particles but not high enough to represent the deep UV limit. It is an ideal value to choose in order to see both the formation and the relatively fast decay of unstable particles due to their release into the colder environment of the bath. Thus much of our discussion will centre around values of $\kappa$ not too distant from this.
\begin{figure}[h!]
\begin{center}
\includegraphics[width=16cm]{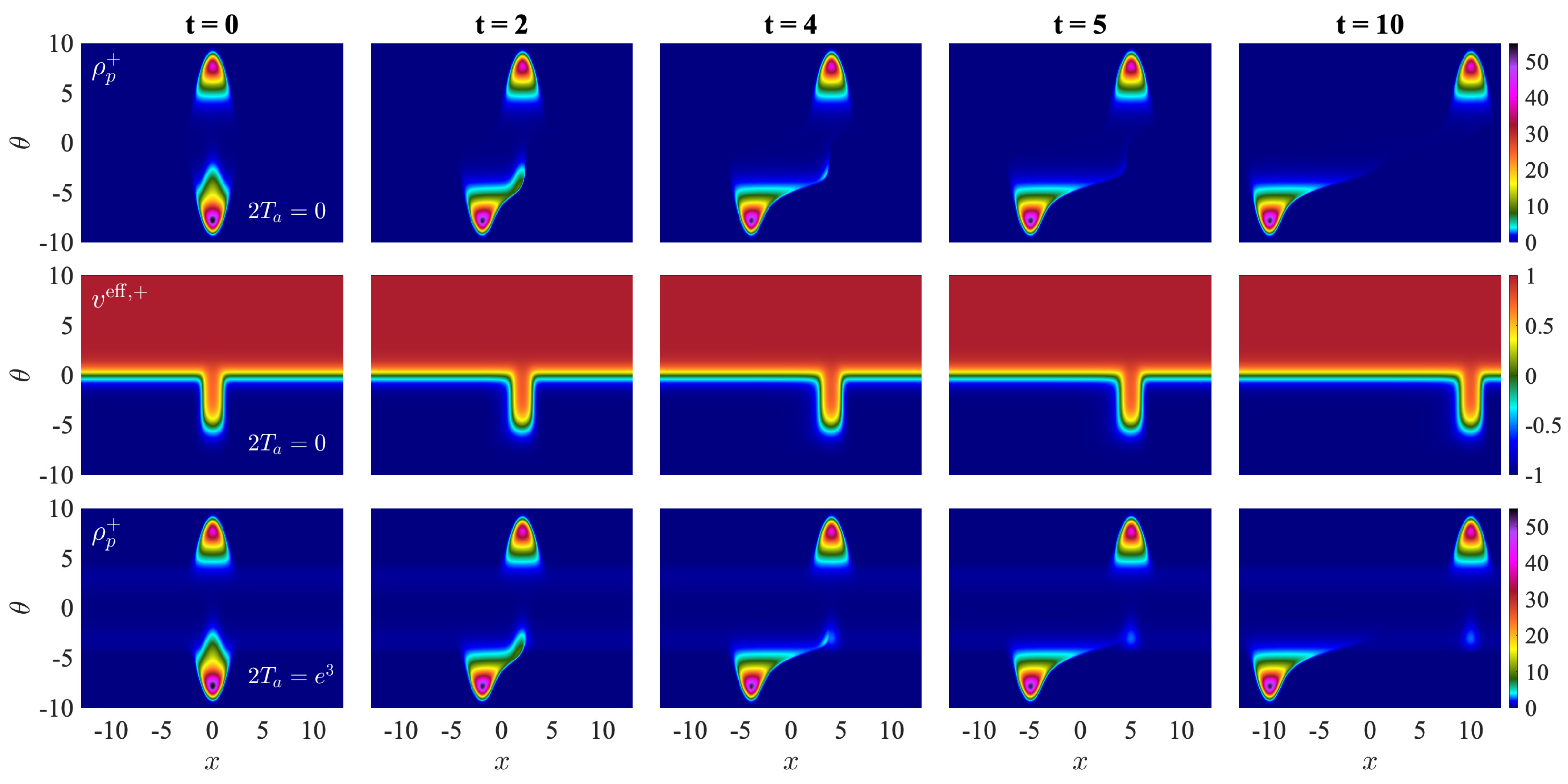}
\end{center}
\caption{Spectral density $\rho_p^+(x,t,\theta)$ in the absence $T_a=0$ (1st row) and in the presence $\log(2T_a)=3$ (3rd row) of a bath for the initial condition (\ref{gaussian}). In both cases $\log(2T_m)=7.5$ and $\sigma=12$ so that $\kappa=1.5$. The three local maxima of the spectral density are clearly visible, particularly for $t\geq 4$ when they are all well separated. They are centered around $\theta=\pm \log(2T_m)=\pm 7.5$ and $\theta=\log(2T_m)-\sigma=-5.5$. The smallest maximum (subsidiary peak) is seen to fully decay into a tail in the first row, whereas it partially persists in the third row. The second row shows the effective velocity for the case of no bath (it is nearly identical with a bath). For fixed $x$, the velocity profiles are mostly $\tanh\theta$-like, except for a small region of values around the subsidiary peak, where the generic $\tanh$-shape is shifted by $-\sigma/2=-5$. Between these two regimes the effective velocity develops intermediate plateaus.} 
\label{rho1}
\end{figure}

The spectral particle density $\rho_p^+(x,t,\theta)$ displays three local maxima: a free fermion peak (so-called because it is identical to the free-fermion solution for $\theta>0$ and this is a consequence of the structure of the kernel (\ref{sphases})) which propagates with speed $+1$ and is centered at $\theta\approx\log(2T_m)$; an interacting peak that propagates with speed $-1$  an is centered at $\theta\approx-\log(2T_m)$ and a subsidiary peak centered at $\theta\approx \log(2T_m)-\sigma$ which contains particles with a range of effective velocities. The values around which the peaks centre in rapidity space are exact for thermal equilibrium with temperature $T_m$, but only approximate in our case. We will see in Section \ref{thetas} how their position depends non trivially for instance on $T_a$ as well. 
In the absence of a bath, the subsidiary peak decays in time leaving behind a tail of slow particles while in its presence it partially decays for early times and partially persists for intermediate and late times. We will study this subsidiary peak in a lot of detail in subsection \ref{subrho}. At the same time, the effective velocity in the absence of a bath, has a shape which is very nearly like the free solution $\tanh\theta$ for most values of $x$, but changes (i.e. experiences mainly a shift of $-\sigma/2$ in rapidity space) around values of $x$ coinciding with the position of the subsidiary peak -- and stays of this form even after the decay of this peak. These three local maxima can be seen in Fig.~\ref{rho1} both without (row 1) and with a bath (row 3), and the effective velocity in the absence of a bath is shown in row 2, providing a visual recap of the main results of \cite{tails}. 

\begin{figure}
\begin{center}
\includegraphics[width=16cm]{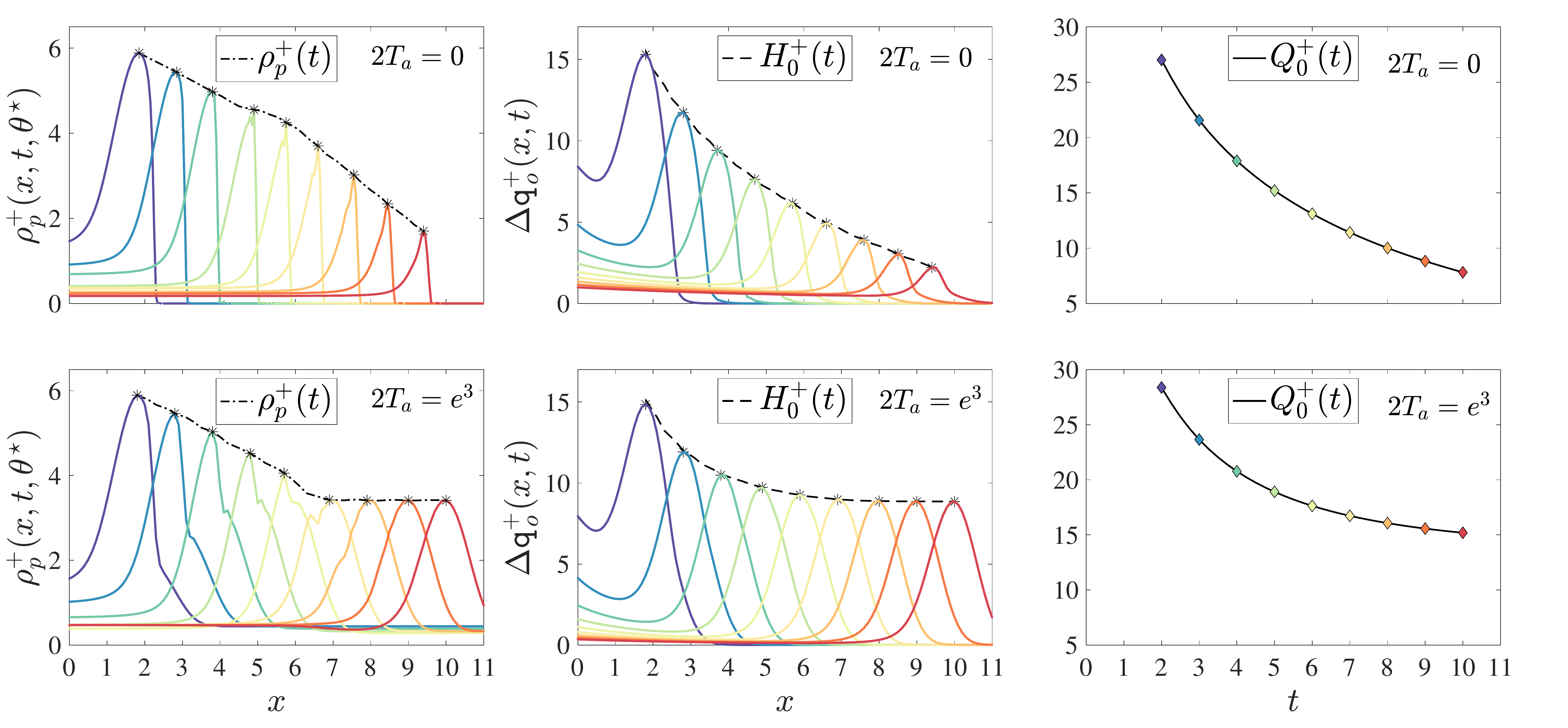}
\end{center}
\caption{Left: The function $\rho_p^+(x,t,\theta^\star)$ and its maximum (stars) $\rho^+_p(t)$. Middle: The function 
$\Delta \texttt{q}_0^+(x,t)$ and its maximum (stars) $H^+_0(t)$. Right: The function $Q^+_0(t)$. In all cases $\sigma=10$, $\log 2 T_m =7$. The top (bottom) row corresponds to the situation without (with) a bath and times are $t=2,3,\ldots,10$. 
}
 \label{decpersist}
\end{figure}

It is the subsidiary decaying (or persisting) peak that is of central interest to us, as it encapsulates the most interesting phenomenology of the problem, namely, that which is related to the creation and decay of unstable particles. Most of this paper will be dealing with the dynamics of this peak.  For a given time $t$ the position of the maximum of this peak is $(x^\star(t),t,\theta^\star(t))$. For short, we will call the value of the spectral density at this point 
\beq
\rho_p^+(t):=\rho_p^+(x^\star(t),t,\theta^\star(t))\,.
\label{rhop}
\eeq
A similar function can be defined for particle ($-$), which is related to the above by a parity transformation:
\beq
\rho_p^-(t):=\rho_p^+(-x^\star(t),t,-\theta^\star(t))\,,
\eeq 
and will indeed be identical to $\rho_p^+(t)$, as the subsidiary peaks are identical for $\pm$ but located at different points in phase space.
In the next section we will investigate some properties of the function $\rho_p^+(t)$. Decay of this function (see left figure, top row, Fig.~\ref{decpersist}) is a signature of the decay of unstable particles. Stationarity, on the other hand, will be a signature of the persistence of a stable density of unstable particles and the emergence of a new magnetic effect (see left figure, bottom row, Fig.~\ref{decpersist}).

A decaying/persisting part is also present in the particle density $\texttt{q}_0^+(x,t)$ but is visually less evident. This is because after integration over rapidities the contributions of the free fermion and subsidiary peak are added up. However, a much clearer picture emerges if the free fermion solution is subtracted.
Let  
\beq
\Delta\texttt{q}_0^{\pm}(x,t):=\texttt{q}_0^{\pm}(x,t)-\texttt{q}_0^{\rm FF}(x,t)\,,
\eeq
be the particle density in the HSG-model minus the free fermion solution defined in (\ref{FFq0}) below. In the absence of a bath, this function has a clear decaying peak, as can be seen in Fig.~\ref{decpersist}, middle image, top row, whereas a persisting peak remains in the presence of a bath as seen in the middle image of the bottom row of Fig.~\ref{decpersist}. Both peaks are essentially the $\theta$-integrated versions of the subsidiary peak of the spectral density.

Let $x^*(t)$ be the space coordinate of the maximum of the decaying peak of $\Delta\texttt{q}_0^{+}(x,t)$ for a given time $t$ (notice that this value is generally different from $x^\star(t)$). Then, we can define the functions
\beqa
&& H_0^+(t):={\Delta\texttt{q}_0^{+}(x^*(t),t)}\,,\qquad  \qquad\qquad
Q_0^+(t):=\int_{0}^\infty dx\, \Delta\texttt{q}_0^{+}(x,t)\,.
\label{defqh}
\eeqa
Notice that the same quantities could be defined starting with the function $\Delta \texttt{q}_0^-(x,t)$ as  $H_0^+(t):={\Delta\texttt{q}_0^{-}(-x^*(t),t)}$ and 
$Q_0^+(t):=\int_{-\infty}^0 dx\, \Delta\texttt{q}_0^{-}(x,t)\,$, so although the functions above carry the $+$ index, they actually give information about both particle types. 

Both functions focus on the decaying/persisting peak of the particle density. Whilst $H_0^\pm(t)$ measures the value of the maximum, $Q_0^\pm(t)$ considers the space integration of the peak and part of the tail, hence it contains information about how the subsidiary peak of the spectral density deforms (loosing its original symmetry in $x$) and either fully disintegrates into a tail or partially persists as a function of time. 

Fig.~\ref{decpersist} presents the three functions $\rho_p^\pm(t)$ (first column), $H_0^+(t)$ (middle column) and $Q_0^\pm(t)$ (last column) in the absence (first row) and presence (second row) of a bath, for several values of time. 
As we can see from these figures, all three functions $\rho_p^+(t)$, $H_0^+(t)$ and $Q_0^+(t)$ decay in time for all times in the absence of a bath or decay for intermediate times and then saturate in the presence of a bath. 

In the next sections we will study this decay/persistence in more detail. Our main objective is to quantify the decay rate of each quantity in terms of the parameters $T_m, T_a$ and $\sigma$.

\subsection{Spectral Density's Subsidiary Peak}
\label{subrho}
The function for which our analysis is most direct is the spectral density itself. As we have seen in Fig.~\ref{rho1}, the function has three local maxima. Let $(\hat{x}(t),t,\hat{\theta}(t))$ be the coordinates of one such maximum, as a function of $t$. Then, we have that
\beq 
\partial_x \rho_p^+(\hat{x}(t),t,\hat{\theta}(t))=\partial_\theta \rho_p^+(\hat{x}(t),t,\hat{\theta}(t))=0\,,
\eeq 
and, consequently
\beq
\partial_t \rho_p^+(\hat{x}(t),t,\hat{\theta}(t))=\frac{d}{dt} \rho_p^+(\hat{x}(t),t,\hat{\theta}(t))\,.
\eeq 
It follows then, that the conservation equation (\ref{cons}) evaluated at the maximum can be written as
\beq
 \frac{d}{dt} \rho_p^+(\hat{x}(t),t,\hat{\theta}(t))=-\Gamma(t) \rho_p^+(\hat{x}(t),t,\hat{\theta}(t))\,\qquad \mathrm{with}\qquad \Gamma(t):=\partial_x v^{\rm eff, +}(\hat{x}(t),t,\hat{\theta}(t))\,.
\eeq 
This implies that the manner in which the value of  the spectral density at its maxima changes is directly  related to the slope of the effective velocity. In particular, the two local maxima associated to the free fermion and interaction peak remain essentially unchanged in time (as we can see in Fig.~\ref{rho1}) and this is consistent with the effective velocity having zero derivative at these points. Indeed, the slope of the effective velocity is only non-trivial around the position of the maximum of the subsidiary peak as we can see in the middle row of Fig.~\ref{rho1}.

\begin{figure}[H]
    \centering
    \includegraphics[scale=0.30]{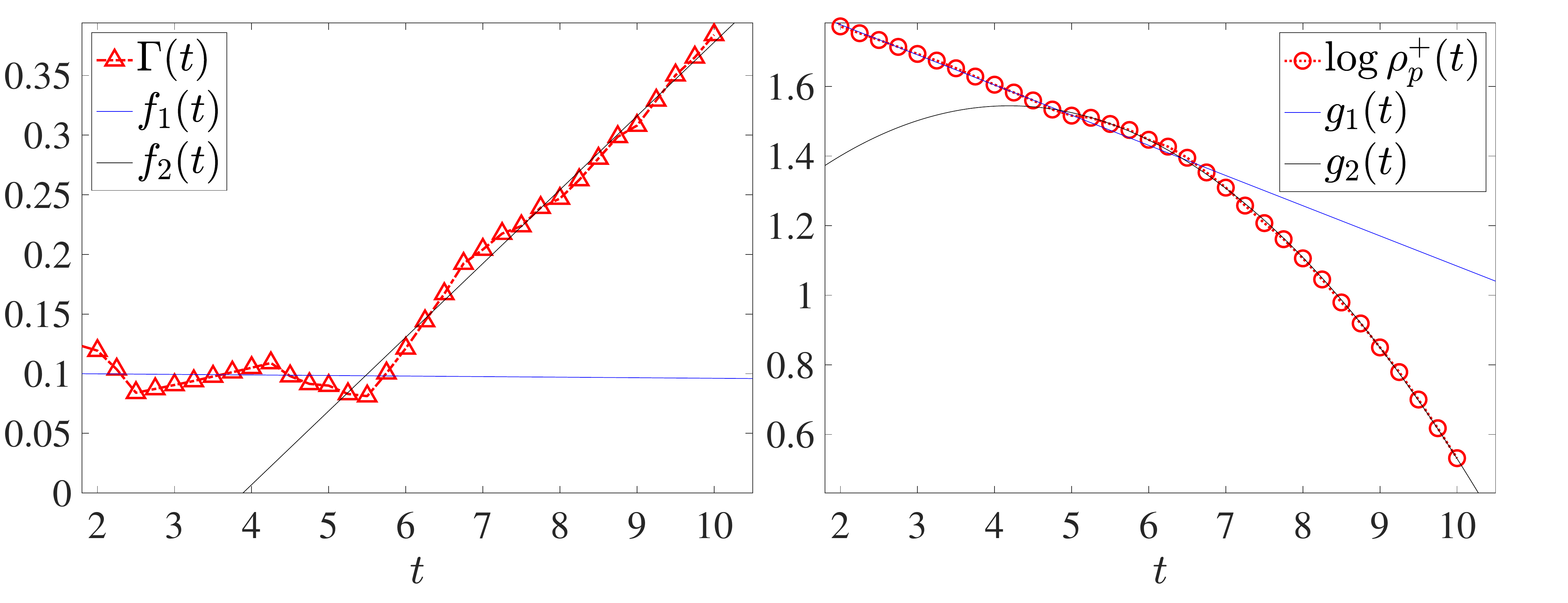}
    \caption{Left: The function $\Gamma(t)$ at the maximum of the subsidiary peak, displaying nearly constant behaviour for intermediate times and linear behaviour for later times. 
    Right: The function $\log(\rho_p(t))$ which is related to $\Gamma(t) $ through (\ref{keye}), displaying the expected linear decay for intermediate times and quadratic decay for later times. The functions $f_1, f_2, g_1$ and  $g_2$ are given in Table~\ref{table:fit_decay} and are mutually related as expected from (\ref{keye}). In both pictures the simulation parameters are $\sigma=10$, $\log(2T_m)=7$ and $T_a=0$. }
    \label{fig3}
\end{figure}

For the subsidiary peak, the equation above becomes an equation for the function $\rho_p^+(t)$
\beq
 \frac{d}{dt}\rho_p^+(t)=-\Gamma(t) \rho_p^+(t)\, \quad \mathrm{that\,\, is} \quad \rho_p^+(t)=\exp \left[-\int dt\, \Gamma(t)\right]
\label{keye}
\eeq
defined in (\ref{rhop}) and it is easy to see numerically that for intermediate times (large enough for the subsidiary peak to be well separated) we have
\beq 
\Gamma(t)\approx \mathrm{constant}\,,
\label{pstat1}
\eeq 
whereas for late times (when the peak has almost fully decayed)
\beq 
\Gamma(t)\approx \mathrm{linear}\,,
\label{pstat2}
\eeq
in other words, $\rho_p^+(t)$ decays approximately exponentially for intermediate times and faster (as a gaussian) for late times. 
This is illustrated in Fig.~\ref{fig3} and the accompanying Table~\ref{table:fit_decay}.

Thus for the subsidiary peak, the simple statements (\ref{pstat1})-(\ref{pstat2}) mean that the decay of its maximum as a function of time is due to how the effective velocity 
changes in position space around the maximum. At the same time we know that the non-trivial additional plateau of the effective velocity in rapidity space around the position of the subsidiary peak is a direct consequence of the presence of unstable particles (it is only formed above the threshold for their formation). Thus, through the conservation equations (\ref{cons}) there is a direct, formal connection between the formation and decay of the subsidiary peak and the creation and decay of unstable particles. 

The decay rate of the maximum of the subsidiary peak (what we called $1/d_\Gamma\approx q_\rho$ in Table~\ref{table:fit_decay}) is exactly the inverse of the slope of the effective velocity. This decay occurs whether or not there is a bath, although in the presence of a bath it is masked by the simultaneous persistence of part of the peak.

Exploring further how these fitting parameters change with $T_m, T_a$ and $\sigma$ is (numerically speaking) a challenging task for the function $\rho_p^+(t)$. Instead, we leave this kind of discussion to the next subsection, where a numerical study of the functions $H_0^+(t), Q_0^+(t)$ defined in (\ref{defqh}) is performed. 

\subsection{The Functions $H_0^+(t)$ and $Q_0^+(t)$}
Although this is not immediately obvious from Fig.~\ref{decpersist}, the manner in which the functions $H_0^+(t)$ and $Q_0^+(t)$ decrease is markedly different for early, intermediate and late times. For this reason we will look at these three regimes separately, with particular focus on the intermediate times regime, where exponential decay is found.
\subsubsection{Early Times: Gaussian Decay}
{\begin{figure}[t]
 \begin{center} 
 \includegraphics[width=16.5cm]{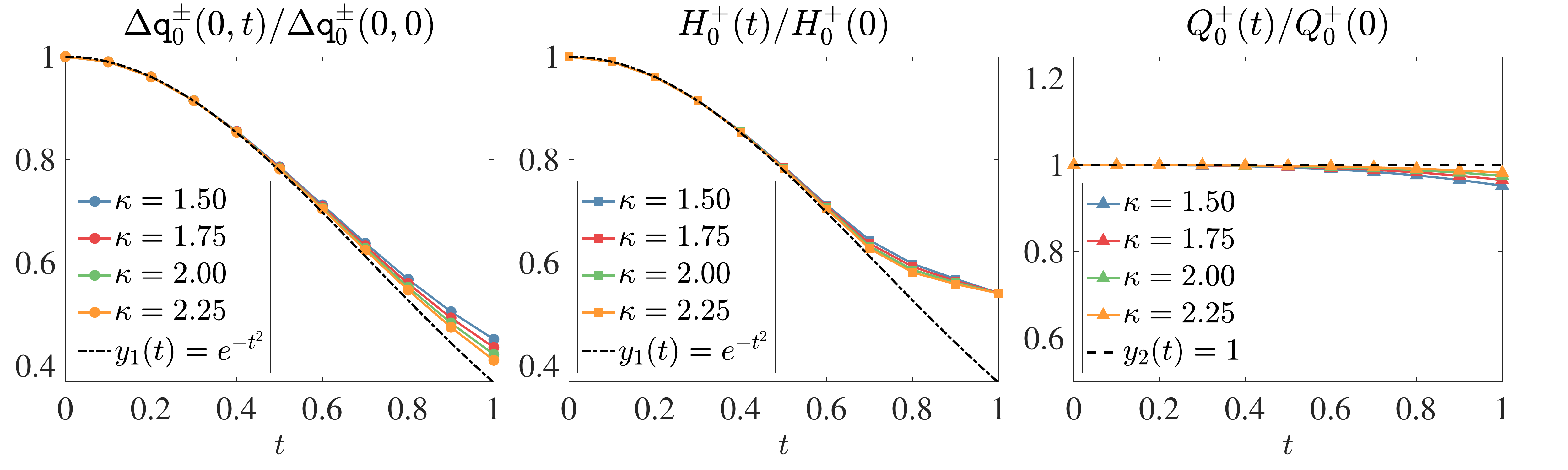}
 \end{center} 
 \caption{Early times behaviour for various values of $\kappa$ and no bath. The function $\Delta \texttt{q}_0(0,t)/\Delta \texttt{q}_0(0,0)$ scales as $e^{-t^2}$, as predicted analytically (\ref{uni}). The same scaling is found for $H_0^+(t)/H_0^+(0)$ for early times, which is expected, as explained below. The range of times for which the gaussian fit works well is roughly $t_g<0.8$ similar to a free fermion. The function $Q_0^+(t)$ on the other hand is nearly time-independent for early times. This is related to the presence of interaction, which for very early times holds the subsidiary peak together, before it starts to decay and loses its symmetry in the spacial coordinate.}  
 \label{earlytime} 
 \end{figure}}
The properties of the functions $H_0^+(t)$ and $Q_0^+(t)$ for early times are fully determined by the initial state temperature profile (\ref{gaussian}). This can be shown analytically using standard equilibrium TBA arguments which are presented in Appendices~\ref{App:A} and \ref{App:B}. 
The first quantity that is helpful to compute is the value of $\texttt{q}_0^+(0,0)$. This is of course an equilibrium quantity, which only depends on $T(0)=T_m$ (\ref{gaussian}). A simple TBA computation allows us to show that for $T_m\gg 1$ 
\beq
\texttt{q}^{\pm}_0(0,0)= \frac{K^\pm(0)}{\pi}\, T_m\,,
\eeq
where $K^\pm(0)$ represent the values of the shifted $L$-functions (\ref{shifted}) at the centre of their plateau in the UV limit. For other parity-symmetry theories the plateau is centered around $\theta=0$, that is $K(0)$ is simply replaced by $L(0)$. For our model, the values of $K^{\pm}(0)$ are known to be \cite{ourtba}
\beq 
K^{\pm}(0)=\log\left(\frac{3+\sqrt{5}}{2}\right)\,.
\eeq
It is also easy to show, by studying the free fermion that the small time behaviour of $\texttt{q}^{\pm}_0(0,t)$ is determined by the value above (for the free fermion $L(0)=\log 2$) and a time dependence which is entirely determined by the gaussian (\ref{gaussian}). This result can be extrapolated to the interacting regime and the initial condition (\ref{gaussian}) to show that
\beq
\texttt{q}_0^\pm(0,t) \approx \texttt{q}^{\pm}_0(0,0) (R+(1-R)e^{-t^2})\qquad \mathrm{for} \qquad R, t \ll 1 \qquad \mathrm{and} \qquad R=\frac{T_a}{T_m}\,.
\label{smallt}
\eeq
Thus, we have that for early times
\beq
\Delta \texttt{q}_0^\pm(0,t) \approx \frac{1}{\pi} \log\left(\frac{3+\sqrt{5}}{4}\right)  (T_a+(T_m-T_a)e^{-t^2})\,.
\eeq
A useful quantity is the normalised function
\beq
\frac{\Delta \texttt{q}_0^\pm(0,t)}{\Delta \texttt{q}_0^\pm(0,0)} \approx  R+(1-R)e^{-t^2}\,,
\label{ratio}
\eeq
which depends only on the ratio $R$. In particular, for $T_a=0$ we have that
\beq
\frac{\Delta \texttt{q}_0^\pm(0,t)}{\Delta \texttt{q}_0^\pm(0,0)} \approx  e^{-t^2}\,,
\label{uni}
\eeq
and therefore, the early times behaviour of this quantity is universal and independent of any parameters. Similar to the free fermion (see Fig.~\ref{freefig}, left) for early times, the local maximum of $\Delta \texttt{q}_0^\pm(x,t)$ is located at $x^*(t)=0$. In fact, we define ``early times" as the interval $t\in [0,t_g]$ where $t_g$ is the largest value of $t$ for which $x^*(t)=0$. As we discuss in Appendix~\ref{App:A}, for the free fermion $t_g=\sqrt{\log{2}}=0.83$. For interacting theories we do not have an analytical prediction for $t_g$, but the results in Fig.~\ref{earlytime} indicate that $t_g$ is roughly of the same order.

Finally, a comment on the early-time features of the functions $H_0^+(t)$ and $Q_0^+(t)$, as can be seen in Fig.~\ref{earlytime}. As we have seen, the early-times dynamics is determined by the initial conditions. In addition, it is to be expected that the behaviour of $H_0^+(t)/H_0^+(0)$ for early times is roughly the same as for  $\Delta \texttt{q}_0^\pm(0,t)/\Delta \texttt{q}_0(0,0)$, simply because for early times at $x=0$ the value of the scaled particle density coincides with that of its local maximum, as the maximum is roughly at $x=0$ too. This is no longer the case for later times and the two quantities start to differ. 
As for the function $Q_0^+(t)$, its behaviour is also linked to the initial conditions, which in this case imply that $Q_0^+(t)/Q_0^+(0)$ remains approximately unchanged for early times. This  behaviour can be qualitatively understood by recalling the relationship between $Q_0^+(t)$ and the particle current. This follows from the conservation equation $\partial_t \texttt{q}_0+\partial_x \texttt{j}_0=0$. For early times, the $\pm$ contributions to currents and densities are identical, so that the equation above becomes a conservation equation for $\pm$ contributions, separately. After integration in half space we have $
\partial_t Q_0^+(t) =\texttt{j}^+_0(0,t)$.
For early times after the quench the particle ($+$) net current at the origin of space remains zero for some time as particles travelling towards the right and the left balance each other. This is a consequence of the symmetry of the initial state which accordingly makes  $Q_0^+(t) \approx Q_0^+(0)$ at early times. Eventually the particle densities become asymmetric as particles interact and form tails. Particles of type ($+$) then slow down as an effect of the interaction, and this induces a non-zero net current at the origin with most ($+$) particles propagating leftwards. Thus $Q_0^+(t)$ starts decaying, signaling particle loss in the region $x>0$.

\subsection{Intermediate Times: Exponential Decay}

\begin{figure}[h!]
 \begin{center} 
 \includegraphics[width=1\textwidth]{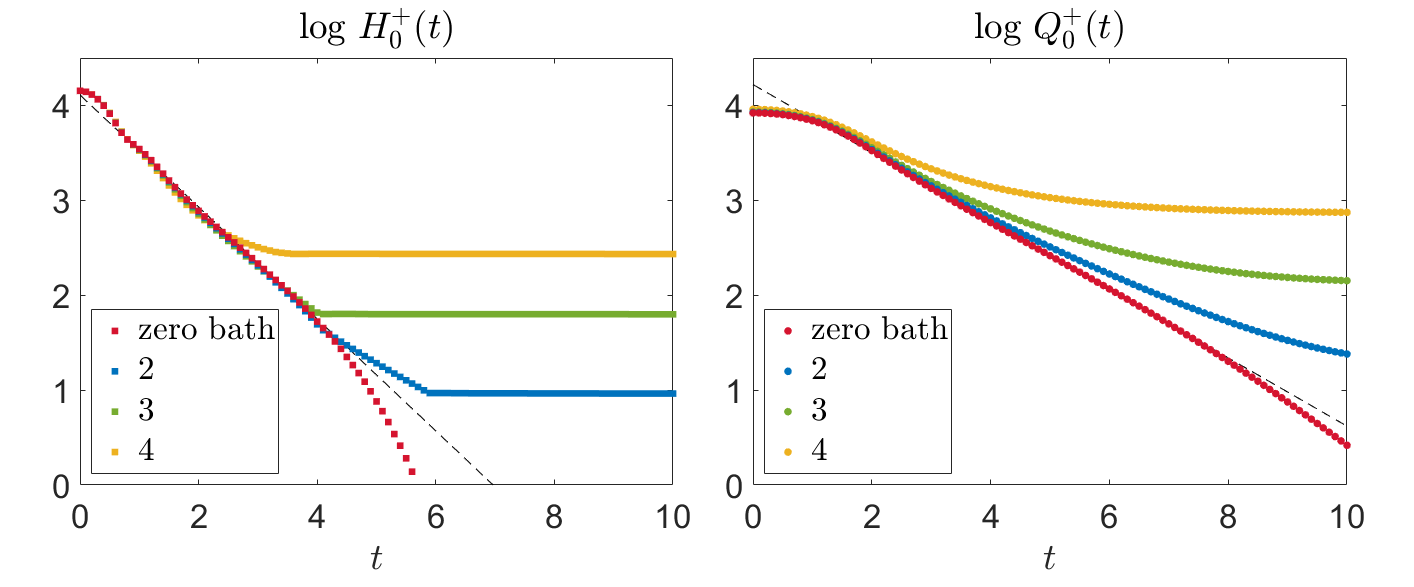}
 \end{center} 
 \caption{Left: The functions $\log{H^+_0 (t)}$  and $\log {Q^+_0 (t)}$ for $\log (2T_m) = 7.5$ and $\sigma =12$ and different bath temperatures $T_a=0$ (no bath) and $\log(2T_a)= 2,3,4$. After an initial transient region, scaling is linear, with slopes that are different for each function and independent of the bath's temperature. The presence of the bath determines only the value of time for which linear decay ends and saturation starts.  We do not report the equations of the fits here as these will be discussed in more detail in Fig.~\ref{samekappa}.}  
 \label{comHQ} 
 \end{figure}

Less trivial is the behaviour for intermediate and large values of $t$. What intermediate values means is strongly correlated with the value of $\kappa$ and also depends slightly on whether we look at $H_0^+(t)$ or $Q_0^+(t)$. We can see the results in Fig.~\ref{comHQ}.

There is a rather large region where the logarithms of both functions behave linearly in time.  For equal parameters, the slopes of the two functions are different from each other. Differences in the decay rates and decay times, are down to the two quantities accounting for different parts of the peaks: $H_0^+(t)$ describes only the maximum, which decays faster, while $Q_0^+(t)$ includes also part of the tail, hence ``capturing" an extra density of particles that have already detached from the subsidiary peak for later times.

\begin{figure}[h]
    \centering
    \setlength{\fboxsep}{0pt}
    \setlength{\fboxrule}{0pt}
          \includegraphics[scale=0.30]{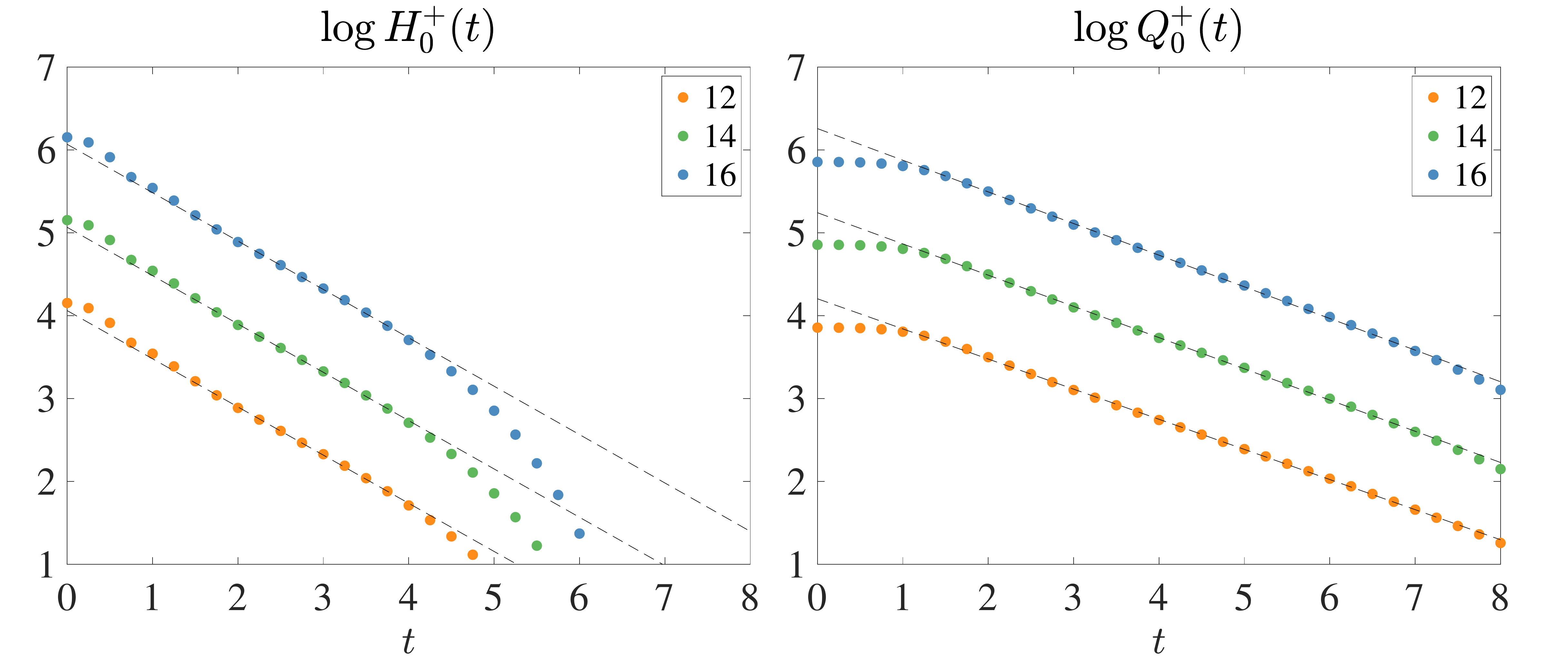}
    \caption{The functions $\log H_0^+(t)$ and $\log Q_0^+(t)$ for fixed $\kappa=1.5$ and various choices of $\sigma=12, 14, 16$. Recall that $\log(2T_m)= \kappa + \frac{\sigma}{2}$. In all cases $T_a=0$.
    The intermediate times linear fit is of the form $\log H_0^+(t)= -\frac{t}{\tau_H} + \xi_H$ and $\log Q_0^+(t)= -\frac{t}{\tau_Q} + \xi_Q$  where $\tau_H\approx1.7$ and $\tau_Q\approx 2.7$. The values of all fitting parameters are summarised in Tables~\ref{tab:kappa15} and \ref{tab:kappa15Q} of Appendix C. For early times $H_0^+(t)$ decays as a gaussian whereas $Q_0^+(t)$ remains constant.
    }
    \label{samekappa}
\end{figure}

Let us now investigate in more detail the parameters upon which the decay rate of functions $H_0^+(t)$ and $Q_0^+(t)$ depends for intermediate times. Fig.~\ref{samekappa} shows very clearly that for sufficiently large $\sigma$ this decay rate is a function of $\kappa$ only. All the figures correspond to the choice $T_a=0$ for which the linear scaling extends over a wide range of times, however recall from Fig.~\ref{comHQ} that the presence of a bath does not modify the decay rate.
\begin{figure}[H]
    \centering
    \includegraphics[scale=0.30]{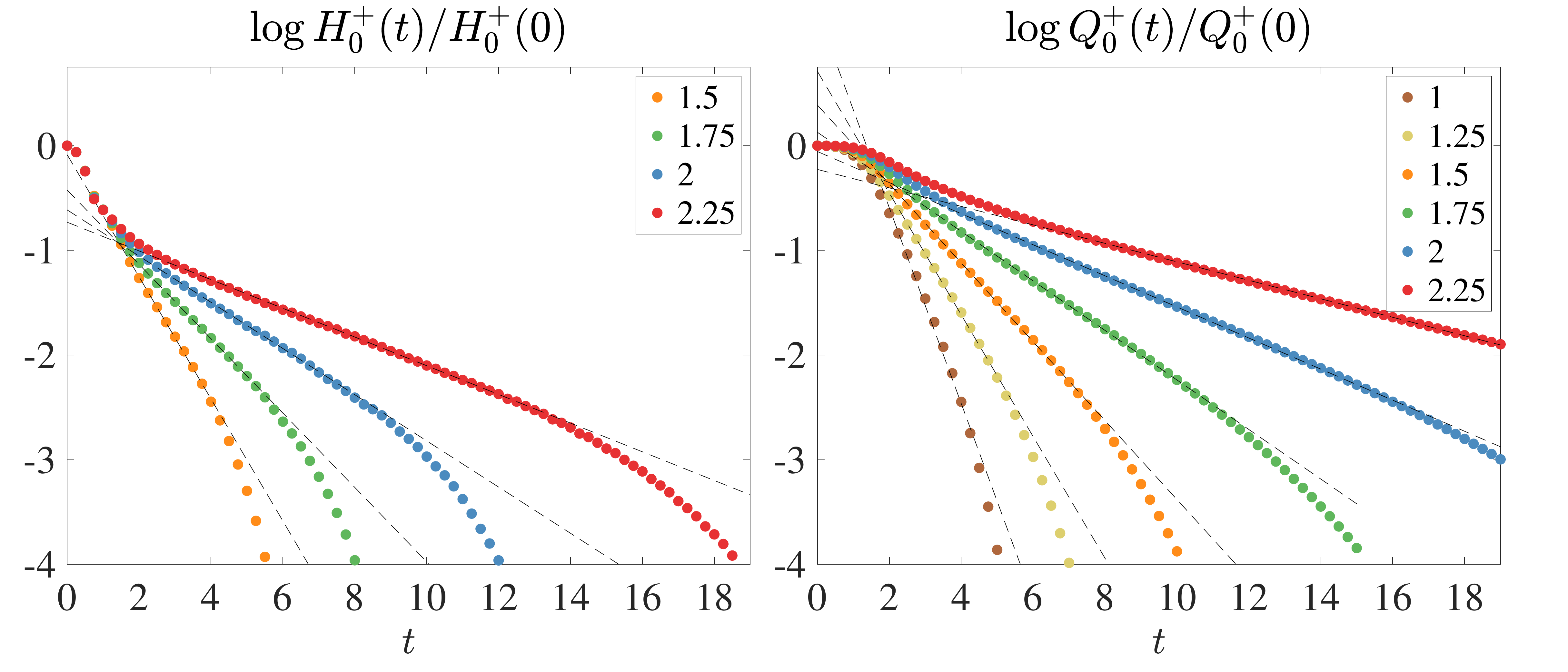}
      \caption{The logarithm of functions $H_0^+(t)$ and $Q^+_0(t)$ normalised by their value at $t=0$ for different values of $\kappa$ in the legends. In the intermediate times region all functions scale linearly with parameters that are summarised in Table~\ref{manyka}, Appendix~\ref{fitt}.
    }
    \label{fig4}
\end{figure}

\begin{figure}[H]
    \centering
     \includegraphics[scale=0.30]{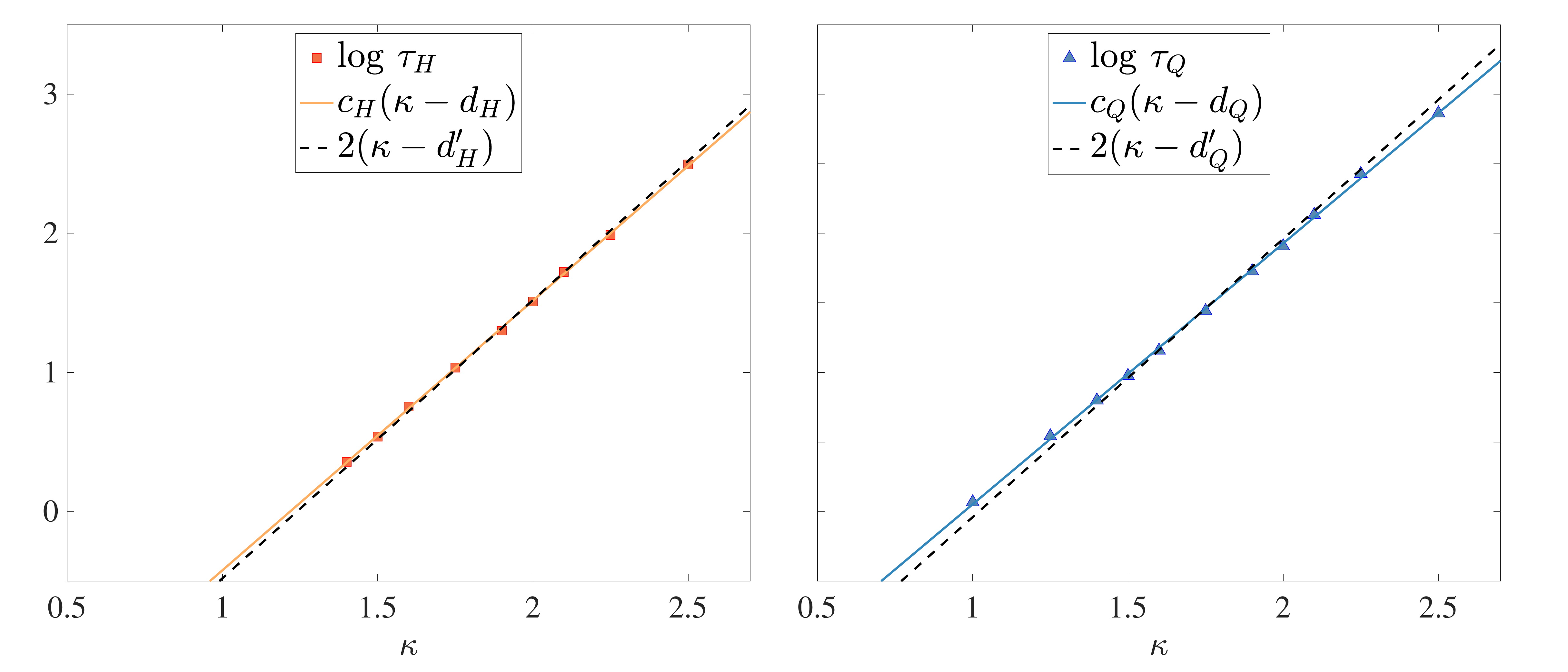}
    \caption{The coefficients $\log \tau_H, \log \tau_Q$ as functions of $\kappa$. The solid lines represent the best linear fits of the points. Fitting parameters are given in Tables~\ref{table4}, Appendix~\ref{fitt} and suggest the functions $\log \tau_H(\kappa)$ and $\log \tau_Q(\kappa)$ are lines of slope $1.94\pm 0.03$ and $1.88\pm 0.03$, respectively. The dashed lines are instead obtained by fixing the slope to 2, and fitting only the intercepts and the resulting coefficients for the two fitting functions $2(\kappa-d^\prime_H)$ and $2(\kappa-d^\prime_Q)$ are $d^\prime_H=1.24\pm0.01$ (with $R^2=0.9986$) and $d^\prime_Q=1.02\pm0.02$ (with $R^2=0.9951$), respectively.}
    \label{fig5}
\end{figure}

Fig.~\ref{fig4} extends the results of Fig.~\ref{comHQ} and Fig.~\ref{samekappa} to further values of $\kappa$, showing that the same type of linear scaling is found for a range of values of $\kappa \in [1.5, 2.25]$ in the interacting regime. For these values the decay of unstable particles is seen most clearly as it happens within reasonably short time-scales. At the same time, in the absence of a bath, the linearity region grows with $\kappa$. This is physically sensible as the more energy there is the system, that is, the higher $\kappa$, the higher the initial particle density and therefore the longer it takes for all these particles to decay.
Table~\ref{manyka}, Appendix~\ref{fitt}, shows the fitting parameters.

Finally, having established that the decay rates of $H_0^+(t)$ and $Q_0^+(t)$, which we will call $\tau_H, \tau_Q$, are functions of $\kappa$ only, it is interesting to investigate what their precise dependence on $\kappa$ is. This is studied in Fig.~\ref{fig5} by fitting the decay rates from Table~\ref{manyka}, Appendix~\ref{fitt}, as functions of $\kappa$. As we can see, the logarithm of both parameters is very well fitted by a line.

In Fig.~\ref{fig5} two possible fits are presented where either both the slope and intercept of the lines are fitted or the slope is fixed to 2 and only the intercept is fitted. The main conclusions are:
\bi 
\item For $\kappa>1$, in the absence of a bath, there is a range of times for which both functions $H_0^+(t)$ and $Q_0^+(t)$ decay exponentially in time. This range of times depends on functions and parameters, but can be roughly defined as the interval bounded by the end of rapid gaussian decay characteristic of early times and the full decay of the peak for late times.
\item The precise scaling is of the form
\beq
H_0^+(t)= h_0 \, e^{-\frac{t}{\tau_H}}\qquad 
Q_0^+(t)= q_0 \, e^{-\frac{t}{\tau_Q}}\,,
\eeq 
where $h_0, q_0$ are constants and 
\beq
\tau_H= e^{c_H\kappa}= \left(\frac{2 T_m}{M}\right)^{c_H}\,, \qquad 
\tau_Q= e^{c_Q\kappa}= \left(\frac{2 T_m}{M}\right)^{c_Q}\,,
\label{newde}
\eeq
that is, within the numerical errors, the decay rates are powers of the universal ratio $T_m/M$ where $M=e^{\frac{\sigma}{2}}$ is the mass of the unstable particle as given in (\ref{tres}).
The precise powers can be read off from Table~\ref{table4}, Appendix~\ref{fitt}, and are 
$c_H=1.94\pm 0.03$ and $c_Q=1.88 \pm 0.03$, that is both very close to the value $2$. Even though $c_Q$ is not compatible with the integer power 2, it is worth pointing out that the quantity $Q_0^+(t)$ is subject to more ambiguities/error sources than $H_0^+(t)$ as it involves an integral and there is some discretion in the choice of integration limits which will introduce an additional error source which is not accounted for in the error reported above. We believe that when this is accounted for, the conjecture $c_H=c_Q=2$ is highly plausible. 
As expected, the time required for a given density of unstable particles to decay grows exponentially with the scale $\kappa$, that is, it is larger the closer the system is to its UV fixed point. 
\ei 
In other words, we can say that for a given time in the interval described above, there exists a stable density of unstable particles that could itself be seen as an unstable excitation in an effective theory. The decay rate of this new unstable excitation takes the universal form (\ref{newde}).

\section{Emergence of a Persistent Peak: a Magnetic Fluid-Like Effect}
\label{magnet}
As we have seen, in the presence of a bath the spectral particle density develops a persisting (subsidiary) peak. That is, there is a finite density of particles that is preserved in time and propagates collectively at speed  $+1$  on top of the bath density. This can be explained as the result of the interaction of the bath with the interacting peak of the opposite particle type, which propagates at speed $+1$. We can make a qualitative analogy between this propagating peak that and a magnet: its effect is to ``drag" the subsidiary peak along, preserving its shape. 

Given the above, we expect that the persisting peak is only formed if there is some overlap in $\theta$-space between the subsidiary peak of the spectral density, and the peaks associated with the bath. Since all these peaks are relatively broad, there is a wide range of temperatures for which this condition is met.  

It is interesting to investigate how this persisting peak depends on the parameters of the model. That is, how does its height, volume, and position in rapidity space depend on  $T_m$, $T_a$ and $\sigma$. We do this in the next three subsections.

\subsection{Persistent Peak's Height and Volume}

Data in Tables \ref{tab:s12}-\ref{tab:bath} provide a numerical in-depth study of the dependence of the persistent peak properties on temperature, for fixed $\sigma$. As usual, we discuss only particle ($+$). 
In particular, we are interested on the maximum height of the persistent peak in the spectral density $h_s:=\rho_p^+(t^\star)$, where $t^\star$ here represents a sufficiently large value of time so that the persistent peak is stable. 
We will also consider the bath ridge height $h_b$, and the volume of the persistent peak $V_s$ at large time $t^\star$. This last measure is computed by integrating the function $\Delta\texttt{q}_0^{+}(x,t)$ in the vicinity of the subsidiary peak maximum 
\begin{equation}
    V_s:= \int_{x^\star-\frac{\delta x_{s}}{2}}^{x^\star+\frac{\delta x_{s}}{2}} d x\  \Delta\texttt{q}_0^{+}(x,t^\star)\ ,
    \label{volume}
\end{equation}
where $\delta x_s$ is the width of the subsidiary peak in the space coordinate space. This has been determined on a case by case basis and depends on the initial conditions.
Furthermore, we consider the relationship between the coordinate $\theta^\star$ (which represents the rapidity at the maximum of the persisting peak in the spectral density) and the position of the maximum of the bath ridge $\theta_b$ as well as the value $\theta_{\rm max}:=\theta_{\rm int}-\sigma$, where $\theta_{\rm int}$ is the rapidity of the maximum of the interacting peak of particle ($-$). We know from the definitions (\ref{sphases}) that interaction between the two species is maximal when rapidity differences are $\pm \sigma$. Therefore $\theta_{\rm max}$ is the rapidity that particles of type ($+$) must have in order to interact maximally with particles of type ($-$) at the point where the latter's density is maximal. Throughout this section, we use the notations
\beq
\alpha_m:=\log(2T_m) \quad \mathrm{and} \quad \alpha_a:=\log(2T_a)\ .
\eeq
Some of the data in Tables \ref{tab:s12}-\ref{tab:bath}  are also presented graphically in the next subsections so that correlations between different parameters are easier to appreciate. However there are some observations that can be made by simply looking at the data:
\begin{table}[h!]
\centering       
{\small                                                                 
\begin{tabular}{|c|c||c|c||c|c|c|}     
\cline{1-7}
$\kappa$ & $\alpha_m$ & $h_s$  &$V_s$ &  $\theta_{\rm max}$ & $\theta_b$  &$\theta^\star$\\
\hline                                                            \hline           
                                                
0 & 6.0 & 0.58 & 0.46 &  -5.65 & -3.22 & -3.22  \\   
\hline

0.5 & 6.5 & 0.82 & 1.26 &  -5.15 & -3.22 & -3.22  \\   
\hline

1 & 7.0 & 1.42 & 3.32 &  -4.66 & -3.22 & -3.15  \\   
\hline                                                                            
1.5 & 7.5 & 2.84 & 8.33 &  -4.17 & -3.24 & -2.97  \\   
\hline                                                                            
2 & 8.0 & 5.89 & 19.89 &  -3.71 & -3.22 & -2.78 \\   
\hline

2.5 & 8.5 & 11.99 & 45.00 &  -3.22 & -3.22 & -2.58 \\   
\hline

3 & 9.0 & 23.25 & 97.67 &  -2.70 & -3.22 & -2.39 \\   
\hline 

3.5 & 9.5 & 53.93 & 222.19 &  -2.17 & -3.22 & -1.94 \\   
\hline   

4 & 10.0 & 100.11 & 442.63 &  -1.71 & -3.22 & -1.83 \\   
\hline  

\end{tabular}}   

\caption{Spectral density's persistent peak data for $\sigma=12, \alpha_a=3$. }   \label{tab:s12}               
 \end{table}
\bi 
\item Both the height and volume of the persisting peak increase rapidly as functions of $\alpha_m$ with fixed $\sigma$. This is to be expected as the energy present in the system is higher for higher $\kappa$ and so the density is expected to increase. 

\item The value of $\theta_b$ is strongly dependent on $\alpha_a$ and close in value to it. If $\alpha_a<\frac{\sigma}{2}$, that is, the bath is in the non-interacting regime, $\theta_b$ is nearly independent of $T_m$ and $\sigma$. In other words, the bath behaves almost like a fully independent free fermion solution at temperature $T_a$.
\item Even though at equilibrium $\theta^\star=\theta_{\rm{max}}$ our data in Table~\ref{tab:bath}
clearly show that this is no longer the case out of equilibrium. In fact, $\theta^\star$ now also depends strongly on $\alpha_a$. These features are seen very clearly in Fig.~(\ref{pos1}).

\item The interacting peak of particles of type ($-$) and the persistent peak of particles of type ($+$) do always overlap in real space, but their maxima do not generically coincide and, in rapidity space, their mutual distance is generally different from $\sigma$. In other words, the maxima of the peaks are not interacting maximally with each other. 
\ei 
Having made these general observations, we will now devote the next subsection to investigating more precisely the correlations between some of these parameters.

 \begin{table}[h!]
\centering       
{\small                                                                 
\begin{tabular}{|c|c||c|c||c|c|c|}     
\cline{1-7}
$\kappa$ & $\alpha_m$ & $h_s$  &$V_s$ &  $\theta_{\rm max}$ & $\theta_b$  &$\theta^\star$\\
\hline  
\hline           

0 & 5.0 & 0.57 & 0.48 &  -4.70 & -3.22 & -3.22 \\   
\hline 

0.5 & 5.5 & 0.74 & 1.16 &  -4.20 & -3.24 & -3.11 \\   
\hline 

1 & 6.0 & 1.13 & 2.75 &  -3.71 & -3.22 & -2.91  \\   
\hline

1.5 & 6.5 & 1.91 & 6.20 &  -3.16 & -3.24 & -2.71  \\   
\hline

2 & 7.0 & 3.41 & 13.33 &  -2.66 & -3.22 & -2.39  \\   
\hline 

2.5 & 7.5 & 6.18 & 27.81 &  -2.23 & -3.24 & -1.98  \\   
\hline 

3 & 8.0 & 10.78 & 53.82 &  -1.69 & -3.22 & -1.74 \\   
\hline                                                                                                                                                                                         
\end{tabular}}   

\caption{Spectral density's persistent peak data for $\sigma=10, \alpha_a=3$.}   \label{tab:s10}                 
 \end{table}
 
 \begin{table}[h!]
\centering       
{\small                                                                 
\begin{tabular}{|c||c|c|c||c|c|c|}     
\cline{1-7}
 $\alpha_a$ & $h_b$ & $h_s$ &$V_s$ &  $\theta_{max}$ & $\theta_b$  &$\theta^\star$\\
\hline  
\hline           
0.0 & 0.02 & 0.11 & 0.36 &  -4.17 & -0.05 & -0.52 \\   
\hline 

0.5 & 0.04 & 0.24 & 0.71 &  -4.17 & -0.33 & -0.80 \\   
\hline 

0.75 & 0.05 & 0.33 & 0.95 &  -4.17 & -0.80 & -1.08 \\   
\hline 

1.0 & 0.06 & 0.45 & 1.23 &  -4.17 & -1.27 & -1.27 \\   
\hline 

1.5 & 0.10 & 0.76 & 2.12 &  -4.17 & -1.74 & -1.74 \\   
\hline 

2.0 & 0.16 & 1.24 & 3.47 &  -4.17 & -2.25 & -2.18 \\   
\hline  

2.5 & 0.27 & 1.92 & 5.49 &  -4.17 & -2.76 & -2.58 \\   
\hline 

3.0 & 0.45 & 2.84 &  8.33 &  -4.17 & -3.24 & -2.97 \\   
\hline 

3.5 & 0.74 & 3.96 &  12.15 &  -4.17 & -3.77 & -3.31 \\   
\hline 

4.0 & 1.22 & 5.19 & 16.96 &  -4.17 & -4.23 & -3.68 \\   
\hline 

4.5 & 2.05 & 6.49 & 23.39 &  -4.25 & -4.76 & -4.05 \\   
\hline 

5.0 & 3.52 & 7.77 & 32.51 &  -4.25 & -5.30 & -4.41 \\   
\hline 

5.5 & 6.24 & 9.42 & 44.61 &  -4.25 & -5.73 & -4.94 \\   
\hline 

5.75 & 8.35 & 10.43 & 54.70 &  -4.25 & -5.99 & -5.56 \\   
\hline 
 
\end{tabular}}   

 \caption{Spectral density's persistent peak data for $\sigma=12, \alpha_m=7.5$, that is $\kappa=1.5$.}                  \label{tab:bath}  
 \end{table}

\subsection{Persistent Peak Position as a Function of Temperatures}
\label{thetas}

\begin{figure}[h!]
    \centering
   \includegraphics[width=1\textwidth]{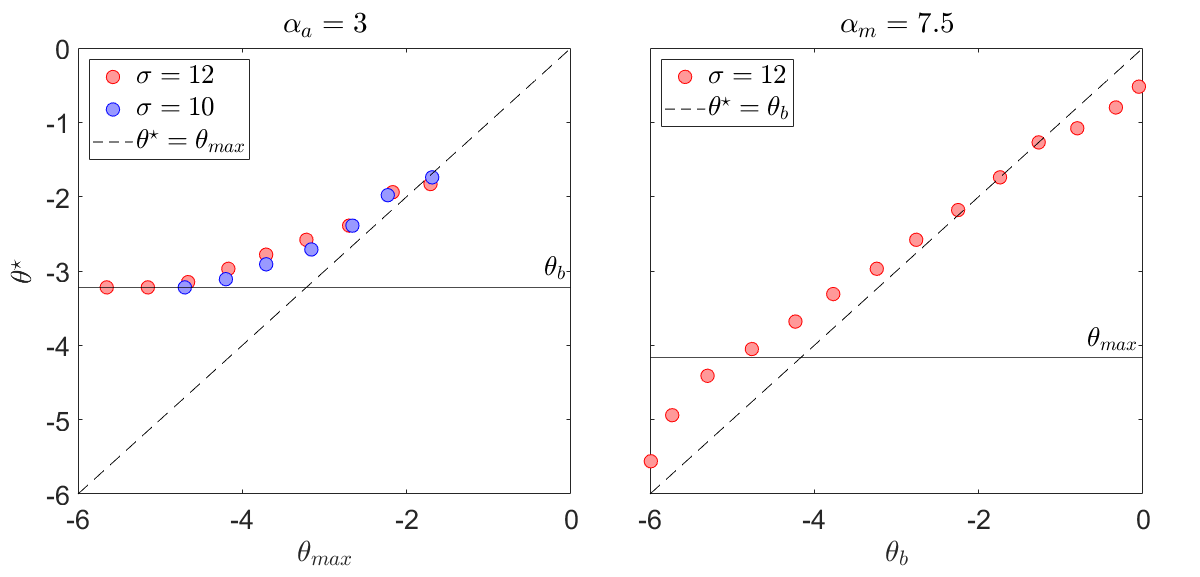}
    \caption{Left: Persistent peak maximum of particle ($+$) position $\theta^\star$ as a function of the rapidity $\theta_{\rm max}$ for which interaction with particles of type ($-$) is maximised. The parameters $\sigma=12, 10$ and $\alpha_a=3$ are fixed. We observe no significant difference for the two values of $\sigma$. The solid black line indicates the bath maximum position $\theta_b$, and the dashed line is $\theta^\star=\theta_{max}$.    
    Right: Persistent peak position $\theta^\star$ as a function of the bath maximum $\theta_{b}$ for a fixed parameter $\sigma=12$ and fixed higher temperature exponent $\alpha_m=7.5$. The solid black line indicates the rapidity maximising the interaction, $\theta_{max}$, and the dashed line is $\theta^\star=\theta_{b}$.    }
    \label{pos1}
\end{figure}

In Fig.~\ref{pos1} we have plotted $\theta^\star$ against
$\theta_{\rm max}$ and $\theta_b$, employing the data in Tables \ref{tab:s12}-\ref{tab:bath}. Our aim is to identify more precisely the relationships between all these parameters. 
Fig.~\ref{pos1} (left) shows visually what we had already observed from the data in the tables of the previous subsection, namely that $\theta^\star$ and $\theta_{\rm max}$ do not generally coincide (they would have done so at thermal equilibrium). Equality is reached only for $\alpha_m$ large provided $\alpha_a$ is in the non-interacting regime (for instance the values $\sigma=10$ $\alpha_m=8$, $\alpha_a=3$ in Table~\ref{tab:s10}). 
For low $\alpha_m$ (though still in the interacting regime), the rapidity $\theta_b$ corresponding to the bath maximum provides a constraint for the position of the interacting peak, as long as the temperature of the bath is large enough to maintain two separate bath ridges (see the right panel in Fig.~\ref{pos1}).
The functional form of the relationship between $\theta^\star$ and $\theta_{\rm max}$ is not clear from the data, but shows
a very slight dependence on the value of $\sigma$, potentially arising only due to the rapidity discretisation.

\begin{figure}[h!]
    \centering
    \setlength{\fboxsep}{0pt}
    \setlength{\fboxrule}{0pt}
    \fbox{\includegraphics[scale=0.48]{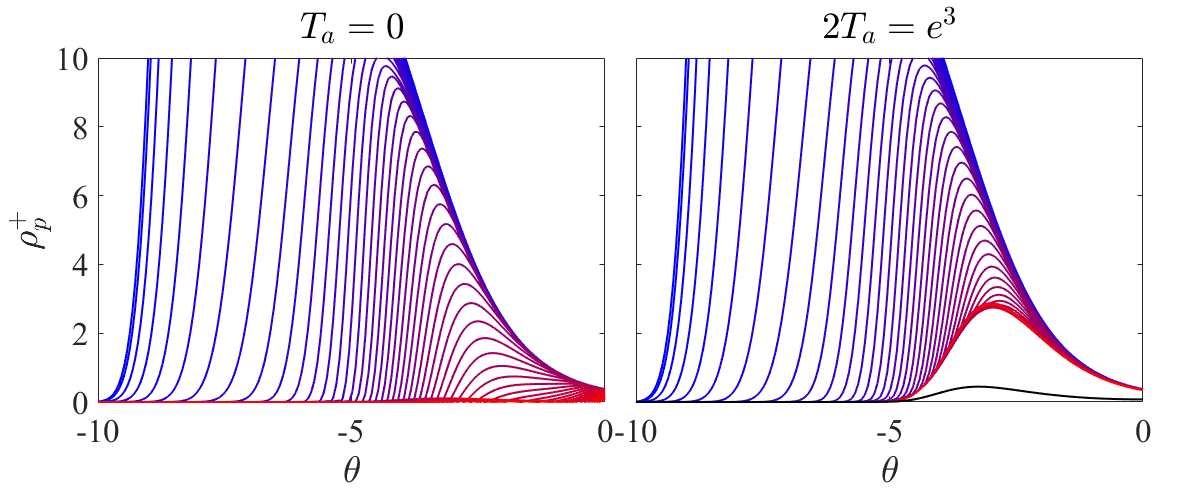}}
    \caption{Cuts in the spectral density $\rho^+_p(t,t,\theta^\star)$ across the subsidiary peak in the plane $x=t$ for times between $0$ and $5$. The parameters are $\sigma=12, \alpha_m=7.5$ and no bath (left) or $\alpha_a=3$ (right). The subsidiary peak decays in the absence of the bath in the left panel and persists taking a shape that ``hugs'' the bath profile plotted in black in the right panel. Note that, as seen in Fig.~\ref{pos1}, right, the maxima do not align.}
    \label{fig:spec_cutsComp}
\end{figure}
Fig.~\ref{pos1} (right) shows the relationship between the maximum of the persisting peak and the maximum of the bath ridge, in rapidity space. We see that these values do not generally coincide either, and their functional relationship is again not obvious. Contrary to the left figure,  this time the parameter $\theta_{max}$, does not serve as a limiting value. This might be partially due to the much larger spread in the rapidity space of the interacting peak.
In rapidity space, the persistent peak maximum is mainly on the right of the bath's maximum, that is in between the two bath ridges. However, the persistent peak ventures beyond the bath maximum for very small values of $\alpha_a$, for which the two bath ridges have merged into one. Note that the bath density is nowhere zero between the ridges for the the temperatures considered here.
This is confirmed also by the spectral density cuts shown in Fig.~\ref{fig:spec_cutsComp} (right). 


 To conclude, the discrepancies between the equilibrium predictions for a Gibbs ensemble at temperature $T_m$ (or $T_a$ for the bath) and the trends we find here are clearly an effect of non-equilibrium time evolution. Interestingly, these deviations do not appear in the scale-invariant quench setup, such as a partitioning protocol \cite{ourUP}. For simplicity, we have considered only the positions of the maxima of peaks. This gives us some useful information but is a gross simplification, as all peaks have different shapes which typically change over time and are asymmetric in rapidity space. In the next subsection we refine our analysis by considering how the height and volume of the persisting peak depend on temperature.

\subsection{Persistent Peak Height and Volume as Functions of Temperature}
\label{hvol}
\begin{figure}[h]
    \centering
    \includegraphics[width=1\textwidth]{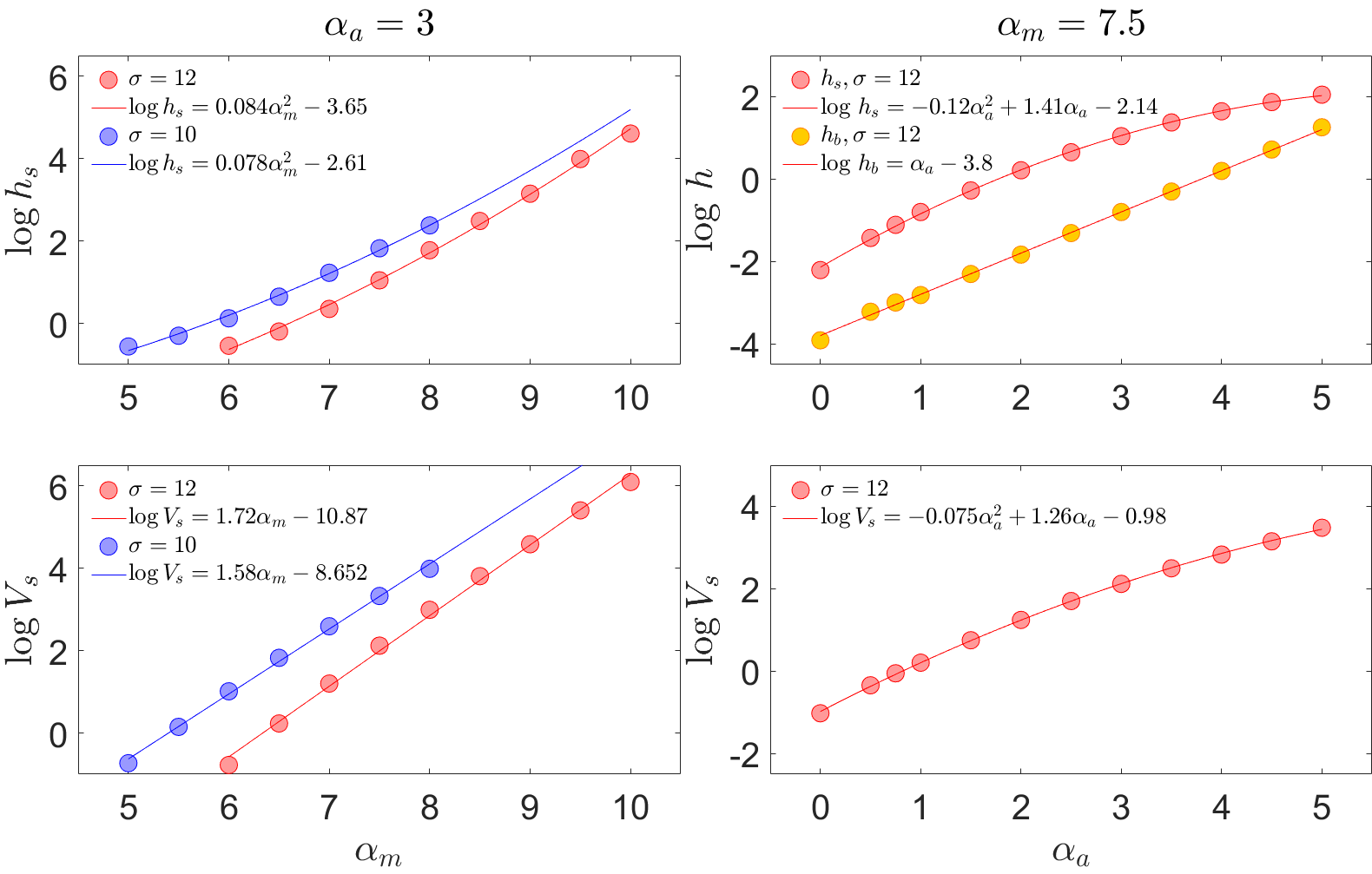}
    \caption{The height and volume of the persistent peak as functions of $\alpha_m$ (left column) and $\alpha_a$ (right column). Note that the fit of $\log h_b$ is in very good agreement with the analytic formula (\ref{fitridge}) in Appendix B. The details of the various fits are shown in Tables \ref{tab:h2}-\ref{tab:V1} of Appendix C.}
    \label{hs_Vs}
\end{figure}

We saw in Tables~\ref{tab:s12}-\ref{tab:bath} that both height and volume grow with the temperatures, with height growing faster. We now quantify this growth in more detail by plotting and fitting the data. 
 Fig.~\ref{hs_Vs}, first column, shows that while the logarithm of the height of the peak increases quadratically with $\alpha_m$, the logarithm of the volume only grows linearly (that is, the volume itself grows linearly with $T_m$). The second column shows that the logarithms of the height and volume of the persisting peak grow quadratically with $\alpha_a$, but unlike in the plots on the left, growth is suppressed for larger values of $\alpha_a$ and the growth rate is generally smaller than with respect to $\alpha_m$.
 
 This gives us an indication that the size of the persistent peak is more strongly dependent on $\alpha_m$ than $\alpha_a$, or in other words, 
 while the persistent peak can only be formed in the presence of an overlapping bath and of an interacting peak of the opposite particle type, the latter has a stronger influence on its height and volume than the former. In fact, for sufficiently large bath temperatures and fixed $T_m$, the growth rate of height and volume of the peak seems almost stalled.
 (see Fig.~\ref{hs_Vs}, right column). 
 Intuitively, the persisting peak will form as long as there is some bath it overlaps with and can propagate over (i.e.~ an available reservoir of particles), but its size is dominantly driven by the strength of the magnet.
 
For the subsidiary peak height $h_s$ the corresponding fits take the form
\begin{equation}
    h_s(\alpha_m)=\exp(a'_m\alpha_m^2+b'_m)\,,
    \label{hsfit}
\end{equation}
with fitting parameters given in Table \ref{tab:h1}, Appendix~\ref{fitt}, and 
\begin{equation}
    h_s(\alpha_a)=\exp(a'_a\alpha_a^2+b'_a\alpha_a+c'_a)\,,
    \label{hsqua}
\end{equation}
with  fitting parameters shown in Table \ref{tab:h2}.
Similarly, for the bath ridge height $h_b$ the corresponding fit takes the form
\begin{equation}
    h_b(\alpha_a)=\exp(\alpha_a+a'_b)\,,
    \label{hbfit}
\end{equation}
with fitting parameters given in Table \ref{tab:hb}, Appendix~\ref{fitt}. The dependence of the bath height on temperature can be derived analytically, as it is almost a free fermion solution at the temperature $T_a$. This calculation is shown in Appendix~B, Subsection~\ref{bath_height} and shows good agreement with the numerical data fit, providing an additional check for the suitability of the rapidity discretisation choice.

For the subsidiary peak volume we find
\begin{equation}
    V_s(\alpha_{m})=\exp(a''_{m}\alpha_{m}+b''_{m})\qquad \mathrm{and}\qquad V_s(\alpha_{a})=\exp(a''_{a}\alpha^2_{a}+b''_{a}\alpha_a+c''_a)\,,
    \label{vollin}
\end{equation}
with fitting parameters presented in Tables~\ref{tab:V1} and \ref{tab:V2}, Appendix~\ref{fitt}, respectively. 


Finally, a word is due on our choice of fittings in Fig.~\ref{hs_Vs}. 
All fits are very good in the sense that they provide the highest accuracy with the smallest number of parameters. However, an analytical derivation or some additional physical intuition, would be desirable. The main achievement of this section is to bring the leading trends to light, that is, to highlight what the relative influences of the two temperatures in the problem on the height and volume of the persisting peak are.

\section{Other Temperature Profiles }
\label{otherthings}
We close this manuscript by briefly exploring other initial conditions defined by temperature profiles in the space coordinate which are distinct from (\ref{gaussian}). We will consider two examples:
\beq
T_{\rm i}(x)=(T_m-T_a)e^{-x^2}+T_a\,, \quad \mathrm{for} \quad T_m, T_a \in \mathbb{R}^+\, \quad \mathrm{and} \quad \sigma> T_a>me^{\frac{\sigma}{2}}>T_m\geq 0\,.
\label{invT}
\eeq 
which we will refer to as ``inverted gaussian profile", that is a profile where the bath temperature is higher than the temperature at $x=0$, and
\beq
T_{\rm d}(x)=(T_m-T_a)\left(\frac{e^{-(x-x_0)^2}+e^{-(x+x_0)^2}}{1+e^{-4x_0^2}}\right)+T_a\,, \quad \mathrm{for} \quad \sigma> T_m>me^{\frac{\sigma}{2}}>T_a\geq 0\,,
\label{doubleT}
\eeq
and $x_0$ is some real value and the normalization is chosen to ensure that $T_{\rm d}(\pm x_0)=T_m$ and $T_{\rm d}(\pm\infty)=T_a$. We call this a ``double gaussian profile".
\begin{figure}[H]
    \centering
    \setlength{\fboxsep}{0pt}
    \setlength{\fboxrule}{0pt}
    \fbox{\includegraphics[scale=0.5]{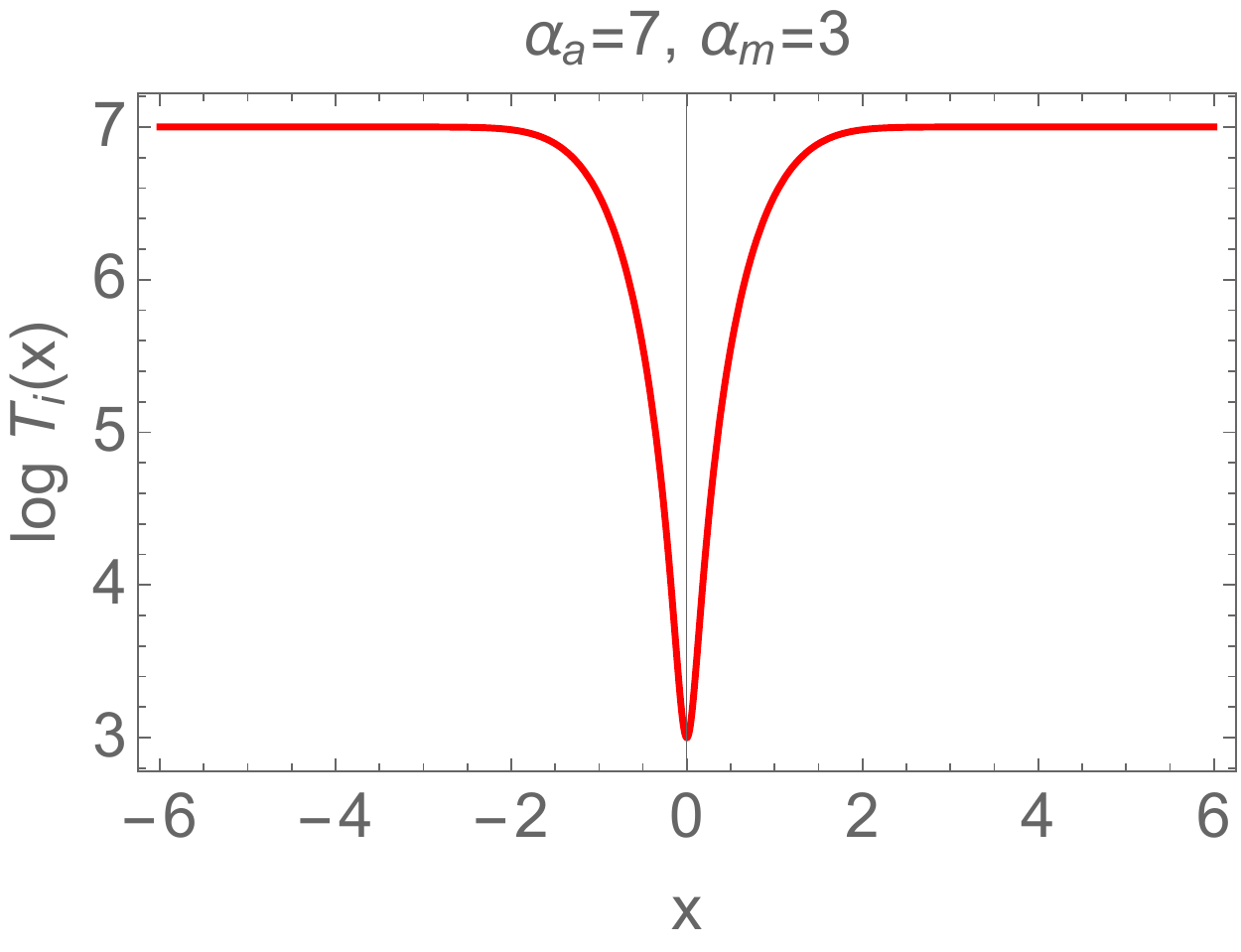}} \qquad 
    \fbox{\includegraphics[scale=0.5]{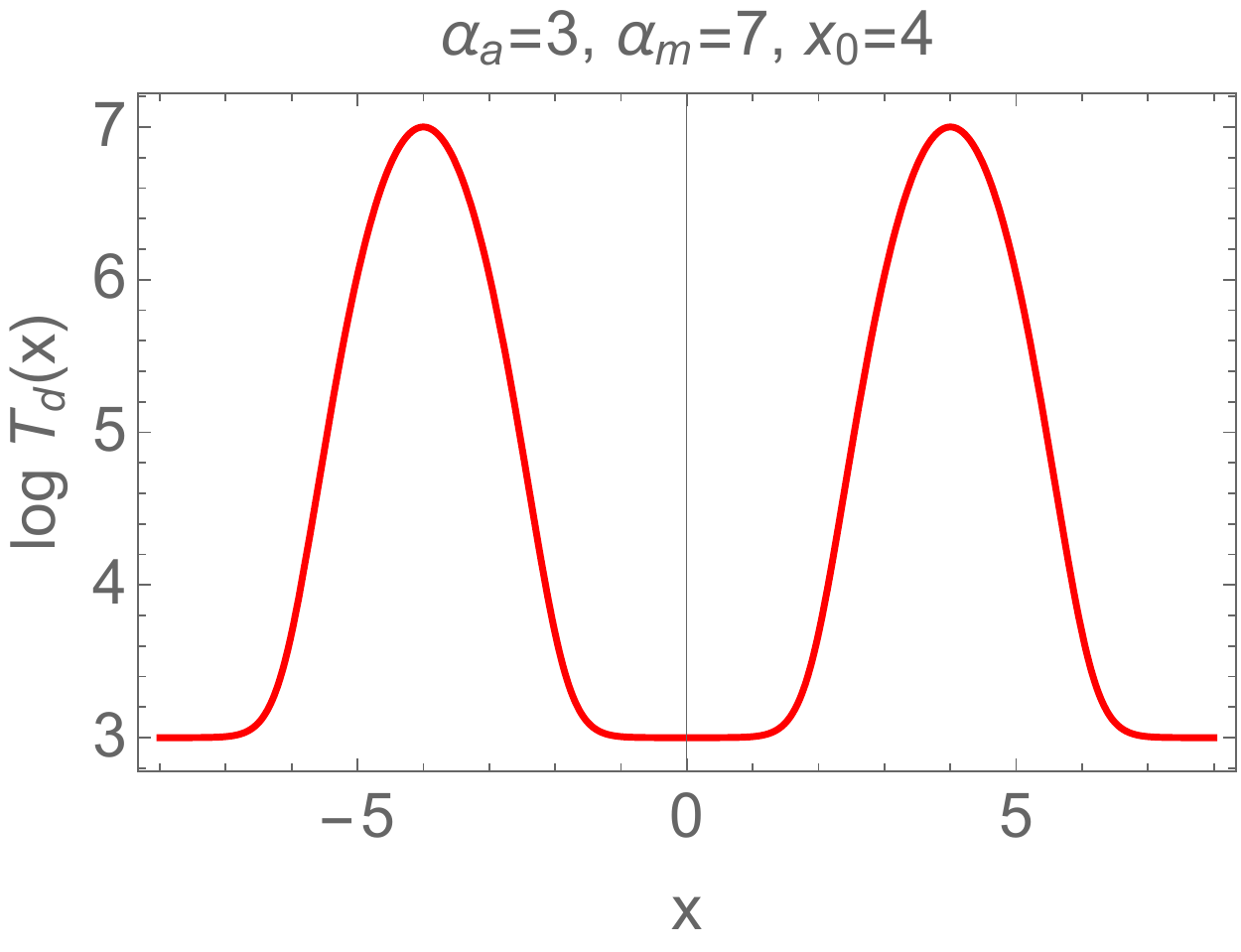}}
    \caption{The inverted gaussian and double gaussian initial conditions for typical values of $\alpha_a, \alpha_m$ and $x_0$.}
\end{figure}
\subsection{Inverted Gaussian Profile:  Going with the Flow}
\label{inverted}
Fig.~\ref{invertedfig} (top row) shows the spectral density $\rho^+_p(x,t,\theta)$ for the initial condition (\ref{invT}) and $\log(2T_a)=7$ and $T_m=0$ with $\sigma=10$.  This gives $\kappa=2$ when defined with respect to the bath temperature. The second row shows the effective velocity and the final row shows the spectral density of a free fermion for the same initial condition (\ref{invT}).

One may have expected that the results would be somehow reminiscent of those obtained for $T_m>T_a$. Instead, they are rather different, as we can see by comparing Figs.~\ref{invertedfig} and \ref{rho1}. The bath region is now at a temperature that allows for the formation of unstable particles whereas a small region around $x=0$ has temperature below the formation threshold. The resulting profiles reflect the fact that the low initial temperatures around $x=0$ are now just a small perturbation of an otherwise high temperature Gibbs ensemble. We can summarise the main features of the spectral density as follows:

\begin{figure}[H]
    \centering
    \includegraphics[width=1\textwidth]{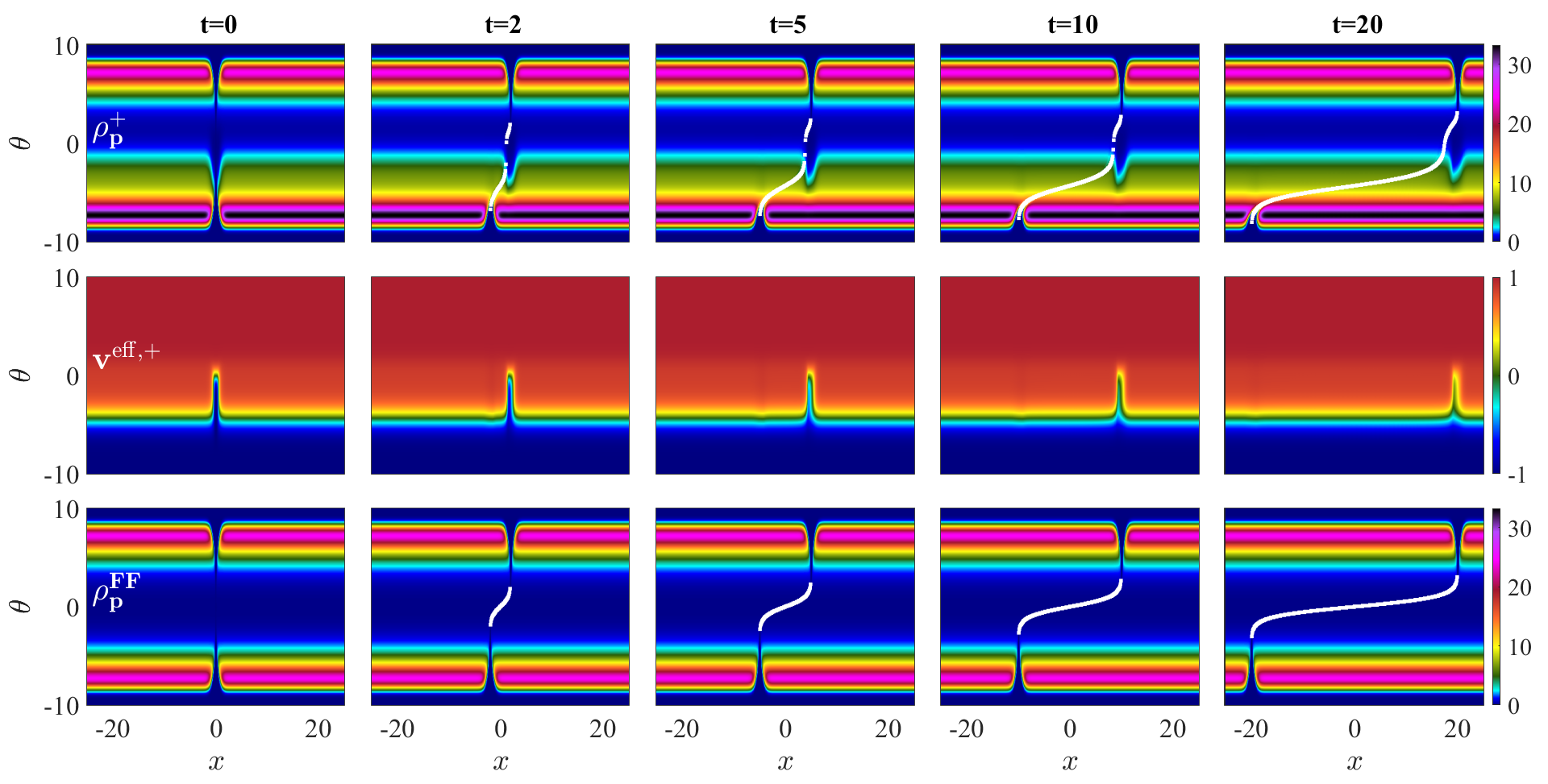} 
    \caption{Spectral density of particle + (row 1) and its effective velocity (row 2) in the case of an inverted gaussian profile with $\log(2T_a)=4$ and $T_m=0$. The white curve superimposed to figures in row 1 is the function $v^{\rm{eff},+}(x,t,\theta)=\frac{x}{t}$ which very accurately describes a narrow zero-density region that is carried by the flow in row 1. Row 3 shows the spectral density of a free fermion for the same initial condition, where the same zero-density filament propagates with the flow and is exactly described by the equation $\tanh\theta=\frac{x}{t}$.}
     \label{invertedfig}
\end{figure}

\begin{figure}[H]
    \centering
    \setlength{\fboxsep}{0pt}
    \setlength{\fboxrule}{0pt}
    \fbox{\includegraphics[width=1\textwidth]{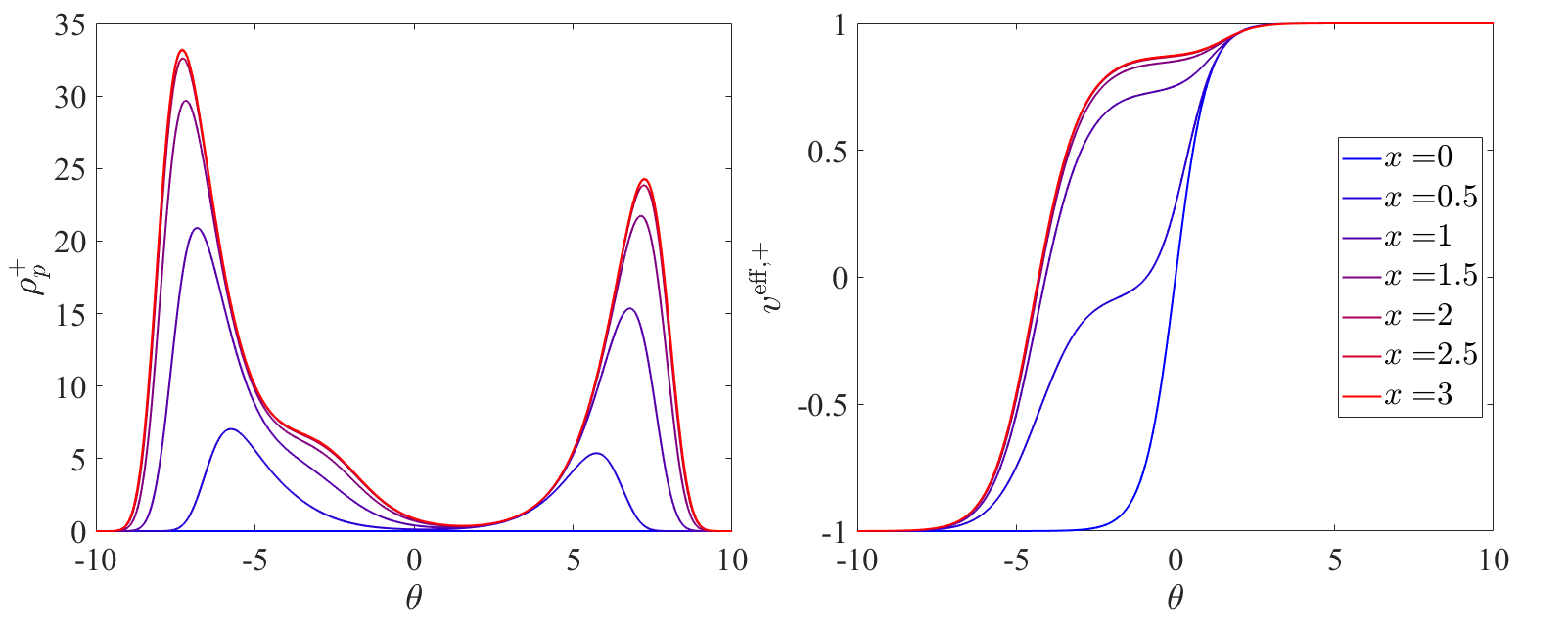}}
    \caption{Cross-sections of the spectral density (left) and effective velocity (right) for the initial condition (\ref{invT}) with $\sigma=10$, $\log(2T_a)=7, T_m=0$. The figures are very similar to those presented in \cite{ourUP} for the thermal equilibrium case, since for this profile most of the system is at a high temperature $T_a$ and resembles a Gibbs ensemble.}
    \label{fig:inverted_eq}
\end{figure}

\bi
\item[(1)] {\bf Equilibrium Features:} At $t=0$ we have that $\kappa=1.5$ for most values of $x$, thus the system is almost entirely in the interacting regime in a Gibbs ensemble. As such most of the features that we see are those of the equilibrium solutions for this temperature as discussed in \cite{ourUP}. This is particularly clear if we look at cross-sections of the spectral density and effective velocity for $x$ fixed, as shown in~\ref{fig:inverted_eq} and compare them to similar figures presented in \cite{ourUP}. Comparing row 2 of Fig.~\ref{rho1} with row 2 of Fig.~\ref{invertedfig} we see that what was a small plateau in Fig.~\ref{rho1} now extends over nearly all values of $x$, except for a small region.

\item[(2)] {\bf Filament Formation} The main novel effect of the low initial temperatures around $x=0$ is the formation of a``filament" corresponding to zero spectral density and which results from the middle region at $x=0, t=0$ being carried by the bath. In other words, the shape of this filament is precisely described by the function $\theta(x,t)$ which is obtained by solving the implicit equation $v^{\rm{eff},+}(x,t,\theta)=\frac{x}{t}$. This curve is superimposed (in white) on the top row of Fig.~\ref{invertedfig}.

\item[(3)] {\bf Free Theory Comparison:} Row 3 of Fig.~\ref{invertedfig} shows the spectral density of a free fermion for the same initial condition (\ref{invT}). As we can see it shares some features with the interacting solution, particularly the formation of a filament of zero density. In the free fermion case its shape is simply described by the function $\tanh\theta=\frac{x}{t}$. Comparing this curve to the function $v^{\rm{eff},+}(x,t,\theta)=\frac{x}{t}$ we see that they are substantially different, especially around $x=t$, another hallmark of the presence of interaction. 

\item[(4)] {\bf Subsidiary ``Dip'':} Row 1 of Fig.~\ref{invertedfig} shows a dip-like feature in the spectral density. Comparing its position with the effective velocity profile in row 2, it is clear that particles surrounding the dip move slower than the observed propagation velocity of $+1$. This is another manifestation of the ``magnetic fluid"-like effect, in which the interacting ridge of particle ($-$) drags along a deformation in the particle ($+$) spectrum. In this sense, the dip in the spectral density is analogous to the subsidiary peak formed in the gaussian temperature profile.
\ei

\subsection{Double Gaussian Profile: Two Persisting Peaks}

Another interesting situation arises when we consider a double gaussian profile.  The main properties of the spectral density can be summarised as follows:

\bi

\item[(1)] {\bf Doubling}: The most obvious feature is the doubling of many of the same patterns found for the profile (\ref{gaussian}), that is, we now have two free fermion peaks, two interacting peaks and two subsidiary peaks that become persistent for large times as well as two tails. We will refer to each of these peaks as right and left, according to their position in real space.

\item[(2)] {\bf New Interactions:} However, the dynamics is more complex than simple doubling of the single gaussian profile features. Time-evolution leads to the merging and/or mutual avoidance of some of these peaks and tails as can be seen in Fig.~\ref{fig:double2} for times larger that 2. A more detailed discussion follows below.

\item[(3)] {\bf Asymptotics:} In the presence of a bath, for large times we recover a familiar picture of persisting peaks riding on bath ridges involving now two peaks, see Fig.~\ref{fig_cutCloseUp}. However, these peaks are not identical due to non-trivial interactions arising for earlier times. 
\ei

\begin{figure}[h!]
    \centering
    \setlength{\fboxsep}{0pt}
    \setlength{\fboxrule}{0pt}
    \fbox{\includegraphics[width=1.0\textwidth]{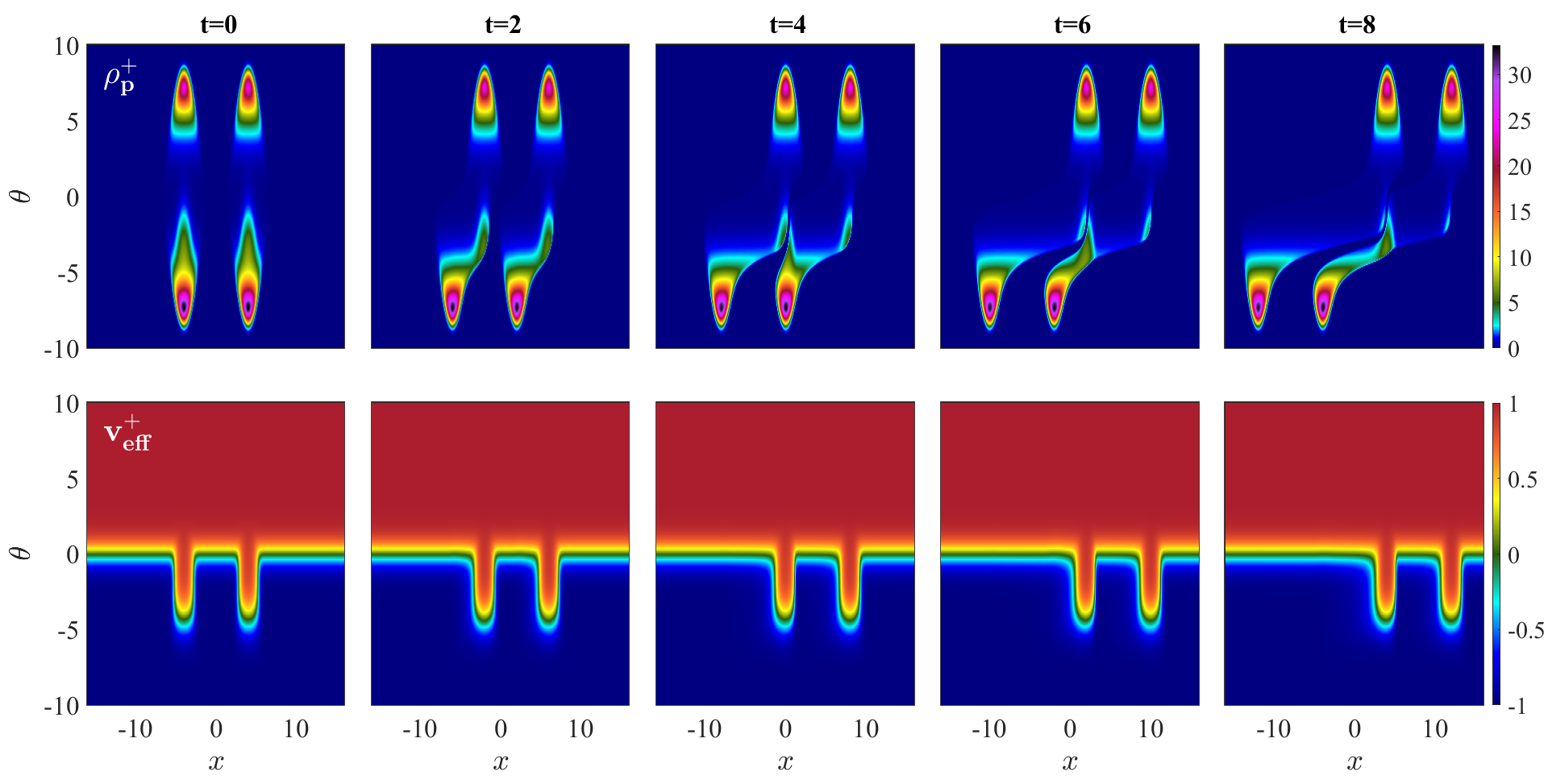}}
    \caption{Spectral density of particle + (row 1) and its effective velocity (row 2) for $t=0$ to $8$ for a double Gaussian profile (\ref{doubleT}). The parameters are chosen here are $\alpha_m=7$, $T_a=0$, $\sigma=10$ and $x_0=4$. The main new feature is that for $t\geq x_0$ the left magnet for particle ($-$) drags some of the particles from the left interacting peak of particle ($+$) into its left subsidiary peak. The later times evolution can be seen in Fig.~\ref{fig:double2}.}
    \label{fig:double1}
\end{figure}

At time $t = x_0 = 4$ in Fig.~\ref{fig:double1}, the left interacting peak of particle ($-$), the left magnet in this scenario, and the right interacting
peak of particle ($+$) are aligned in space. Since their distance in rapidity space is comparable to the
resonance parameter $\sigma$, bound states between some of the particles in these two peaks
are formed. Consequently, the left magnet drags along a matter wave originating from
not only the left subsidiary peak of particle ($+$), but also part of the right interacting peak of particle ($+$). This process starts already at time $t\approx 2$ because of the finite
spread of the peaks in coordinate space.
The additional particles join the left subsidiary peak of particle ($+$) and the right interacting
peak of particle ($+$) becomes smaller and remains smaller in the long-time limit. This
can be visually ascertained by noticing, that the tail of the left interacting peak is
much wider than the tail of the right interacting peak, see Fig.~\ref{fig:double2}. The reversal of the travelling
direction of particles redirected from the interacting peak to the subsidiary peak is accompanied by a negligible change in the velocity profile, hence is purely an interaction effect.
After the interacting peaks of both particle species separate, the two subsidiary peaks of particle ($+$) are dragged by the two magnets of the
($-$) type. Since the left subsidiary peak spectral density was replenished by additional
particles up until the time $t\approx 10$, this is approximately the timescale of the delay
between the decay stages of the right and left subsidiary peaks, as we see in Fig.~\ref{fig:double2}.

Interestingly, Fig.~\ref{fig:double1} also shows a new phenomenon reminiscent of ``tail shyness" that is the fact that the right and left tails of the spectral density appear to consistently avoid each other for all times \cite{wiki}. This is seen more clearly in
Fig.\ref{fig_merge} where various cuts for a fixed rapidity $\theta=-3$ are presented. We observe that the left and right tails avoid touching each other and leave
a finite zero-density region between them. In the magnetic 
fluid picture we can say that although the passing of the type ($-$) magnet leads to a deformation of the spectral density, no particle transfer between the left and right tails ever takes place.
\begin{figure}[h!]
    \centering
    \setlength{\fboxsep}{0pt}
    \setlength{\fboxrule}{0pt}
    \fbox{\includegraphics[width=1\textwidth]{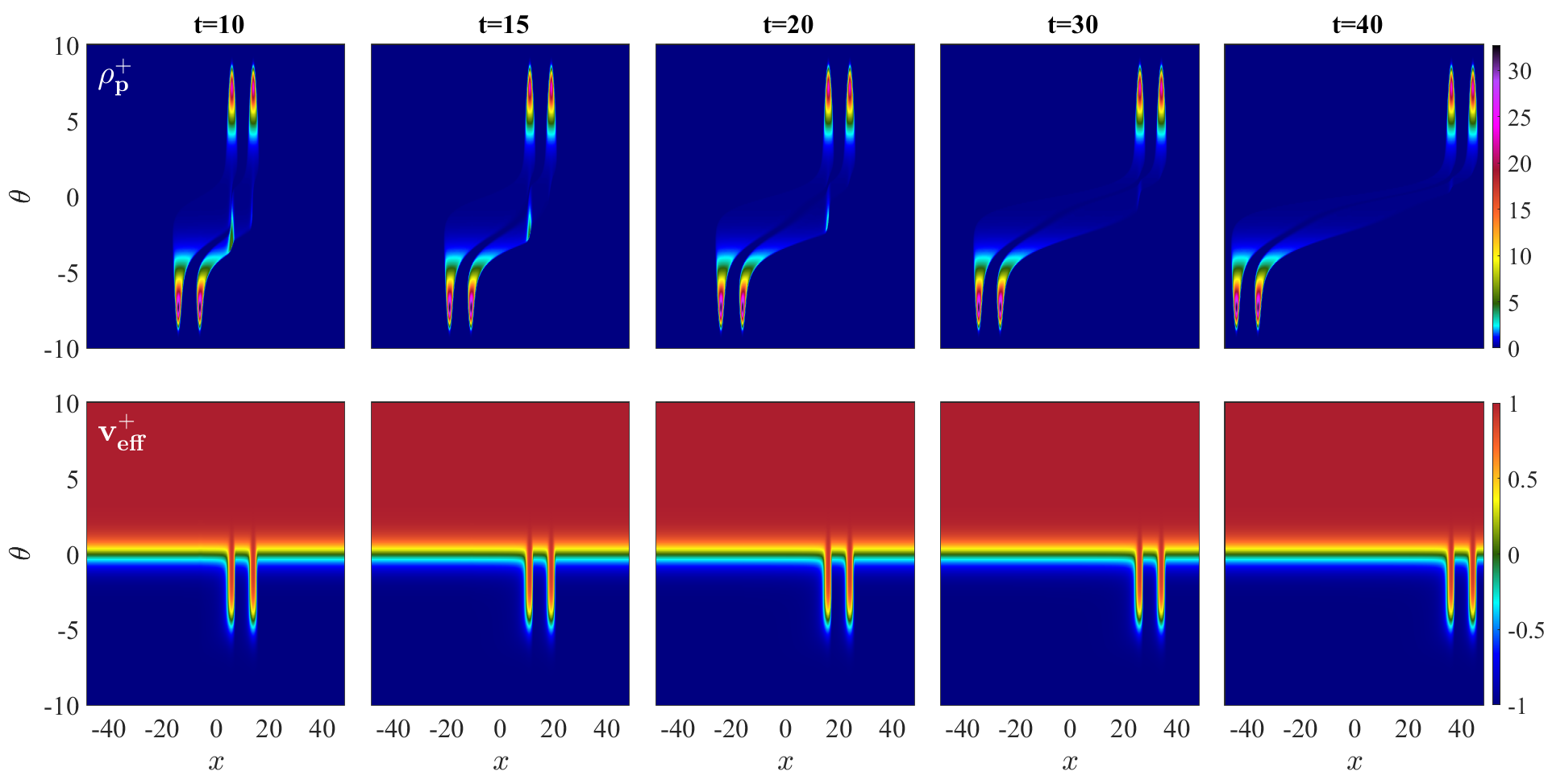}}
    \caption{Spectral density of particle + (row 1) and its effective velocity (row 2) for times $t=10$ to $40$ and double gaussian initial condition. The parameters are chosen here as $\log(2T_m)=7$, $T_a=0$, $\sigma=10$ and $x_0=4$. At late times, the decay of the left subsidiary peak lags behind the decay of the right subsidiary peak by approximately 10 time units.}
    \label{fig:double2}
\end{figure}

\begin{figure}[H]
    \centering
    \setlength{\fboxsep}{0pt}
    \setlength{\fboxrule}{0pt}
    \fbox{\includegraphics[width=1\textwidth]{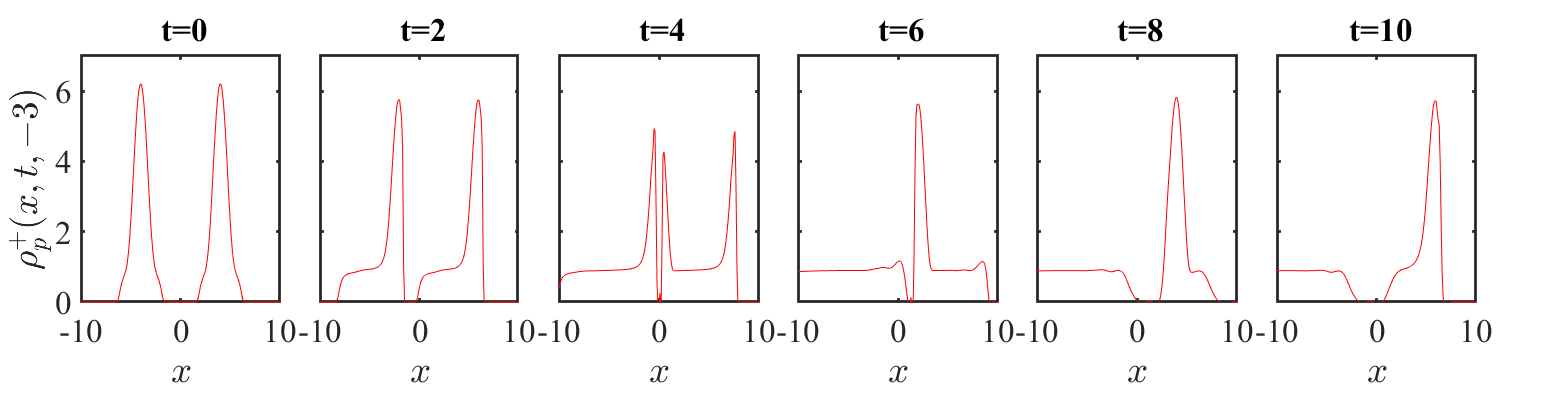}}
    \caption{Spectral density cuts of particle ($+$) for $\theta=-3$ cutting across the subsidiary peaks. Within the time $t=0$ to $10$ we observe the decay of the right subsidiary peak and the left subsidiary peak transfer from the left tail to the right tail. We also see that the tails never touch each other (we call this ``tail shyness" inspired by the phenomenon of crown shyness that is observed in trees \cite{wiki}).}
    \label{fig_merge}
\end{figure}

\begin{figure}[H]
    \centering
    \setlength{\fboxsep}{0pt}
    \setlength{\fboxrule}{0pt}
    \fbox{\includegraphics[width=1\textwidth]{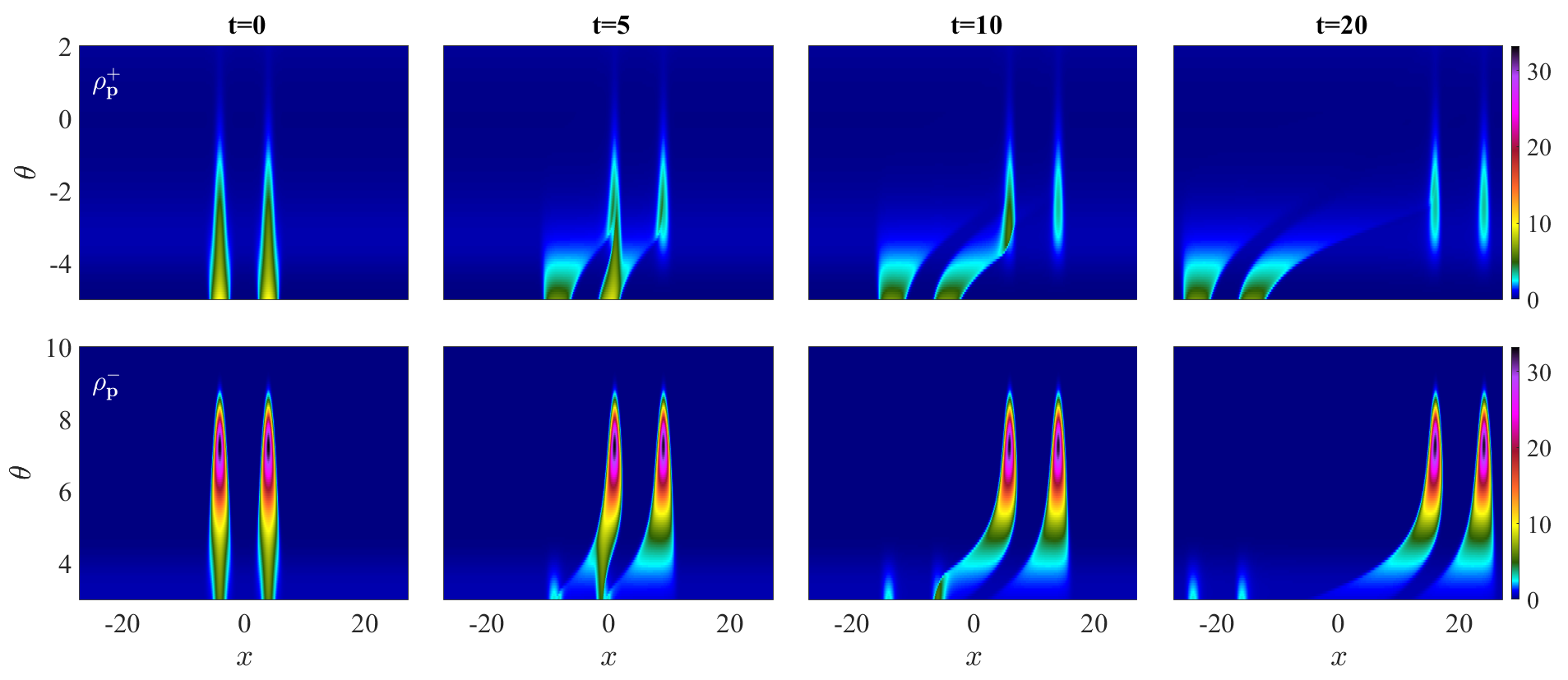}}
    \caption{Spectral density close ups of particle ($+$) (row 1) and particle ($-$) (row 2) for $\sigma=10$, $\alpha_m=7, \alpha_a=3$ and $x_0=4$. $t=0$: Two subsidiary peaks of particle ($+$) and two interacting peaks of particle ($-$) are identical. $t=5$:    
    The left subsidiary peak of particle ($+$) acquires additional particles from the right interacting peak of particle ($+$). $t=10$: The right subsidiary peak of particle ($+$) is persistent whereas the left peak is still decaying. $t=20$: The right persistent peak of particle ($+$) aligns in coordinate space with a larger magnet, the right interacting peak of particle ($-$). As a result, the right persistent peak of particle ($+$) is larger than the left (see the cross-sections shown in the Fig~\ref{fig_cutCloseUp}).}
    \label{fig_closeUp}
\end{figure}

\begin{figure}[h!]
    \centering
    \setlength{\fboxsep}{0pt}
    \setlength{\fboxrule}{0pt}
    \fbox{\includegraphics[width=0.92\textwidth]{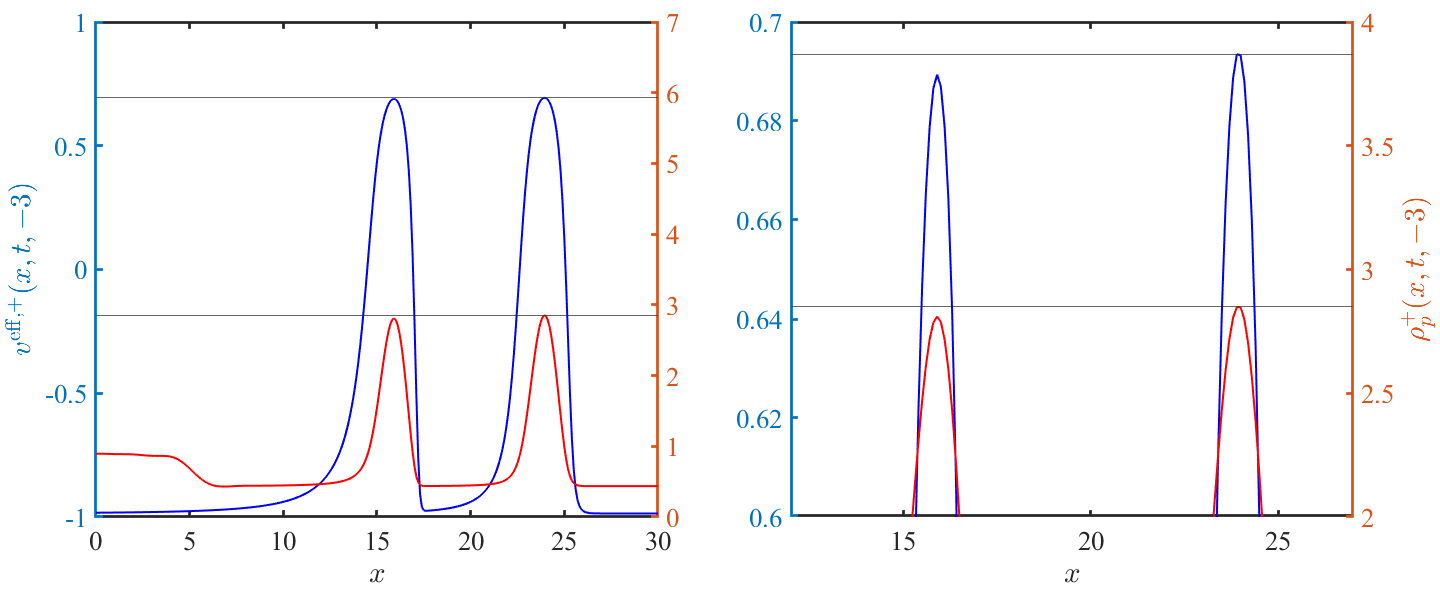}}
    \caption{Spectral density (red) and effective velocity (blue) cuts for $\theta=-3$ and $t=20$. The right is a close-up of the left figure. The right subsidiary peak, which aligns with a larger magnet, is both larger and has a higher effective velocity than the left subsidiary peak, even if their propagation velocities are asymptotically the same. The parameters are $\alpha_m=7$, $\alpha_a=3$, $\sigma=10$ and $x_0=4$.}
    \label{fig_cutCloseUp}
\end{figure}

Another interesting feature are the long-time characteristics of the persistent peaks in the presence of a bath. 
The behaviour in this case is identical to what we have already described for early times, but as expected, for late times decay does not continue. Instead, two persisting peaks are formed. 
Close ups as well as cross-sections of the spectral density are shown in Figs.~\ref{fig_closeUp} and \ref{fig_cutCloseUp}. Although the right subsidiary peak has acquired more matter at early time, in the long time limit the deciding factor is the size of the type ($-$) magnet. The left magnet of particle ($-$) lost some of its particles to a
subsidiary peak of particle ($-$). Therefore, it has fewer available particles to interact
with the subsidiary peak of particle ($+$). As a result, the right magnet interacts with
the bath more strongly and the right persistent peak of particle ($+$) is larger by $\Delta\rho^+_p=0.04\pm 0.02$. Due to
stronger interactions, it also has a higher effective velocity $\Delta v^{\mathrm{eff},+}=0.004\pm 0.002$, although in the end both persistent peaks have a propagation velocity equal to that of the magnets $x^*/t=1$. We expect these differences to be more pronounced for higher temperatures $T_m$.

\section{Conclusion and Outlook}
\label{conclu}
In this paper we have studied the post-quench dynamics of an integrable QFT characterised by two stable particles which interact to form an unstable bound state. The quench under consideration is defined by an initial gaussian temperature profile, with maximum temperature $T_m$ at $x=0$ and a lower temperature $T_a$ or bath temperature for $|x|\gg 1$. 
\medskip 

Our results build on and extend previous work by
 improving the quantitative and qualitative understanding of the main numerical observations of \cite{tails}. These were chiefly that in the presence of one unstable particle and upon release of a high density of its stable constituents into a cold environment, the associated spectral densities will develop tails and a third persistent local maximum (persistent peak). These in turn, translate into an asymmetric particle density profile including also a tail, as well as an additional plateau in the effective velocities. 

A consistent interpretation of these results is that tails are associated with the decay of unstable particles in a cold environment while a persisting peak may be associated with a persisting, stable density of such particles, which can be created thanks to the magnet-like interaction between a low-temperature bath of a certain stable particle type and a high density of the other particle type that the bath bounds with. 

Here we have asked two main questions: if tails are due to particle decay, can be determine the associated decay rate?, and how does a persistent density of particles correlate with the parameters of the theory? We have found detailed answers to both these questions, mainly based on a numerical analysis and occasionally complemented by analytic computations on a free fermion model. 

In order to define a decay rate we first have to choose a suitable quantity or quantities where decay is easy to define. We have been able to find three such quantities: the height of the additional local maximum of the spectral density, $\rho_p^+(t)$, the height of the local maximum of its $\theta$-integrated version (particle density) $H_0^+(t)$, and the $x$-integration of this peak, $Q_0^+(t)$. All of these quantities decrease as functions of time, either for all times (in the absence of a bath) or only for intermediate times (in the presence of a bath). In all cases we have found that whereas for early times, the dynamics is determined by the initial conditions, for intermediate times decay is exponential. Here intermediate times are defined as $t>t_g$, where $t_g$ can be defined as the earliest time for which the typical double-bump split of the particle density occurs. Once this linear region begins it only ends for some late time at which decay either stops (in the presence of a bath) or else, the peak has decayed to the extent that a reliable maximum can no longer be identified.

For the function $\rho_p^+(t)$, the decay rate is simply related to the slope of the effective velocity in real space, as determined by the GHD conservation equation. This has been established both numerically and analytically. For the functions $H_0^+(t)$ and $Q_0^+(t)$ the decay rate is  $\frac{(2T_m)^p}{M^p}$ where $M$ is the mass of the unstable particle and we have conjectured that $p=2$. The dependence on $T_m/M$ only, is strongly supported by numerics, whereas the power $p=2$ includes a numerical error. Thus the decay rate is a function of a universal ratio which compares the overall energy scale (i.e. the highest temperature) with the mass of the unstable particle. The higher this ratio is, the more unstable particles can be formed, and the longer it takes for the same fraction of such particles to decay. 

In the presence of a bath, a persistent peak is formed for a wide range of choices of $T_m, T_a$ and $\sigma$ and both the height and volume of this peak depend on all these parameters. First, for it to be formed at all, we must have $\log(2T_m)>\frac{\sigma}{2}$, that is, there must be interaction present in the system. Once this is guaranteed, then both temperature scales, $T_a$ and $T_m$ have a pronounced influence on the properties of the persisting peak. Whilst the position of its maximum is strongly correlated with $T_a$, its height and volume correlate strongly with $T_m$. 

Finally, we have also explored other initial conditions, namely, an inverted gaussian profile where the temperature in the bath is higher than at $x=0$ and in the interacting regime, and a double gaussian profile, with temperature choices similar to those in the main case. We have identified new phenomena specifically associated with these profiles, notably, in the inverted profile, the low density region around $x=0$ gets carried away by the flow forming a filament whose shape is perfectly described by the function 
$v^{{\rm eff},+}(\theta)=x/t$. For the double gaussian profile we have found instead that the dynamics is not simply two copies of the single-gaussian case but rather the various peaks and tails interact non-trivially, leading to the formation of two distinct tails in the case without bath, and two distinct persistent peaks when a bath is present. 

\medskip 
There are many interesting follow-up problems that one could consider with a view to further exploring the role of unstable particles in the dynamics of integrable systems. Obviously, there is a great variety of quenches that we have not yet  considered and where creation and decay of particles may leave new signatures. Also, the hydrodynamic quantities that we have studied in this paper and in \cite{ourUP, tails} are just the simplest and most accessible. It would be very interesting to investigate for instance how the presence of unstable particles affects the dynamics of two and higher point functions, now also accessible through iFluid, as well as of entanglement.   

Another interesting problem is the study of other theories in the HSG family that posses more than one unstable particle. Not only will this produce new signatures of creation and decay, but could also have some conceptual significance from the viewpoint of scattering theory. Given that, as we have seen, under the right conditions, stable densities of unstable particles can be formed which operate as new effective degrees of freedom, one may wonder whether in the presence of several unstable particles we may even be able to develop an effective scattering theory among these new degrees of freedom. We hope to address these questions in future work. 

\medskip

{\bf \noindent Acknowledgements}: We thank Frederik S. M\o{}ller for discussions on iFluid and for keeping us up to date on improvements and extensions of the program. We are grateful to Alvise Bastianello for stimulating discussions at the early stages of this work.

\appendix
\section{Some Useful Free Fermion Results}
\label{App:A}
Let
\beq
\texttt{q}_0^{\rm FF}(x,t)=\int_{-\infty}^\infty d\theta \rho_p^{\rm FF}(x,t,\theta) = \frac{1}{2\pi}\int_{-\infty}^\infty d\theta\, \frac{\cosh\theta}{1+e^{\beta(x,t,\theta)\cosh\theta}}\,,
\label{FFq0}
\eeq
where 
\beq
\beta(x,t,\theta)=\frac{1}{(T_m-T_a) e^{-(x-t \tanh\theta)^2}+T_a}\,,
\eeq
 be the particle density of a free fermion for the initial state (\ref{gaussian}). Having such an explicit solution makes many analytic computations possible for the free fermion. These provide useful insights into some of the properties of the functions we have investigated in this paper. 
 \subsection{Local Maxima}
\begin{figure}[h!]
    \centering
    \includegraphics[width=0.49\textwidth]{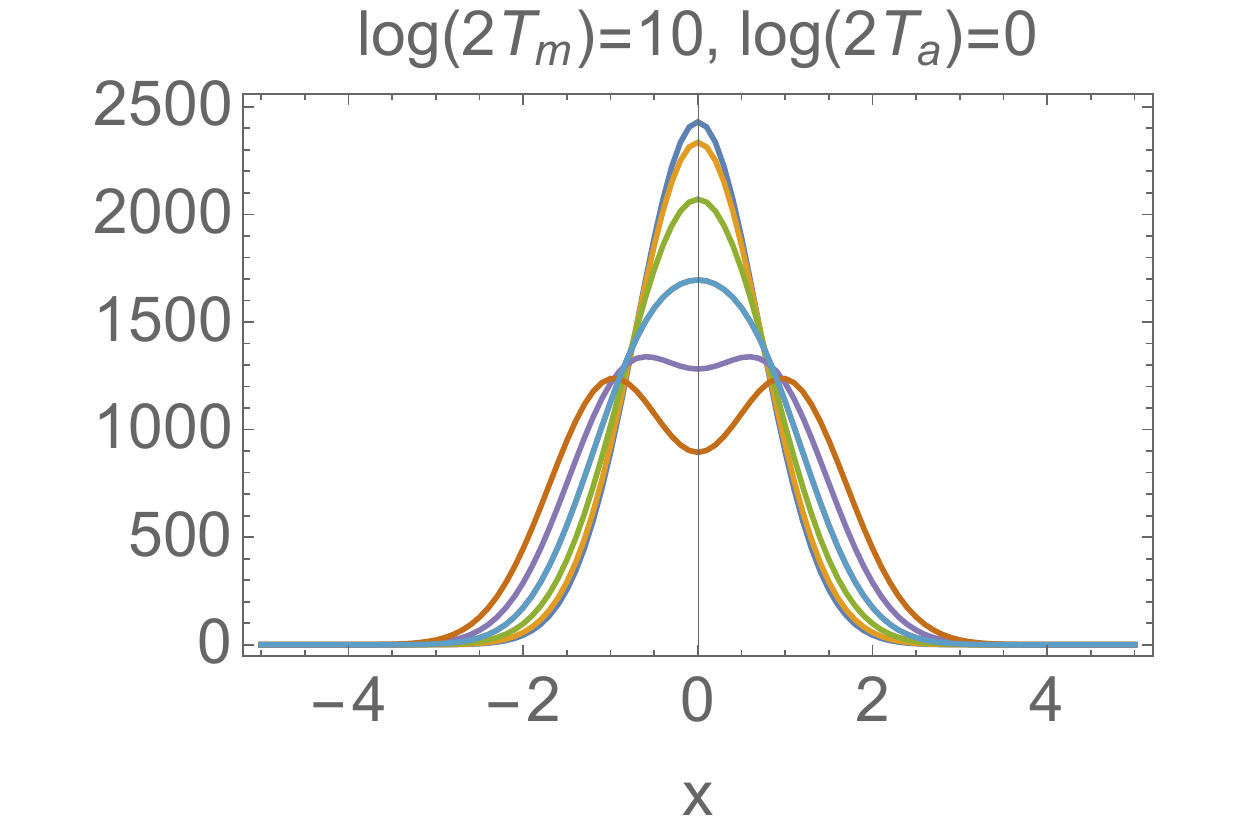}
   \includegraphics[width=0.49\textwidth]{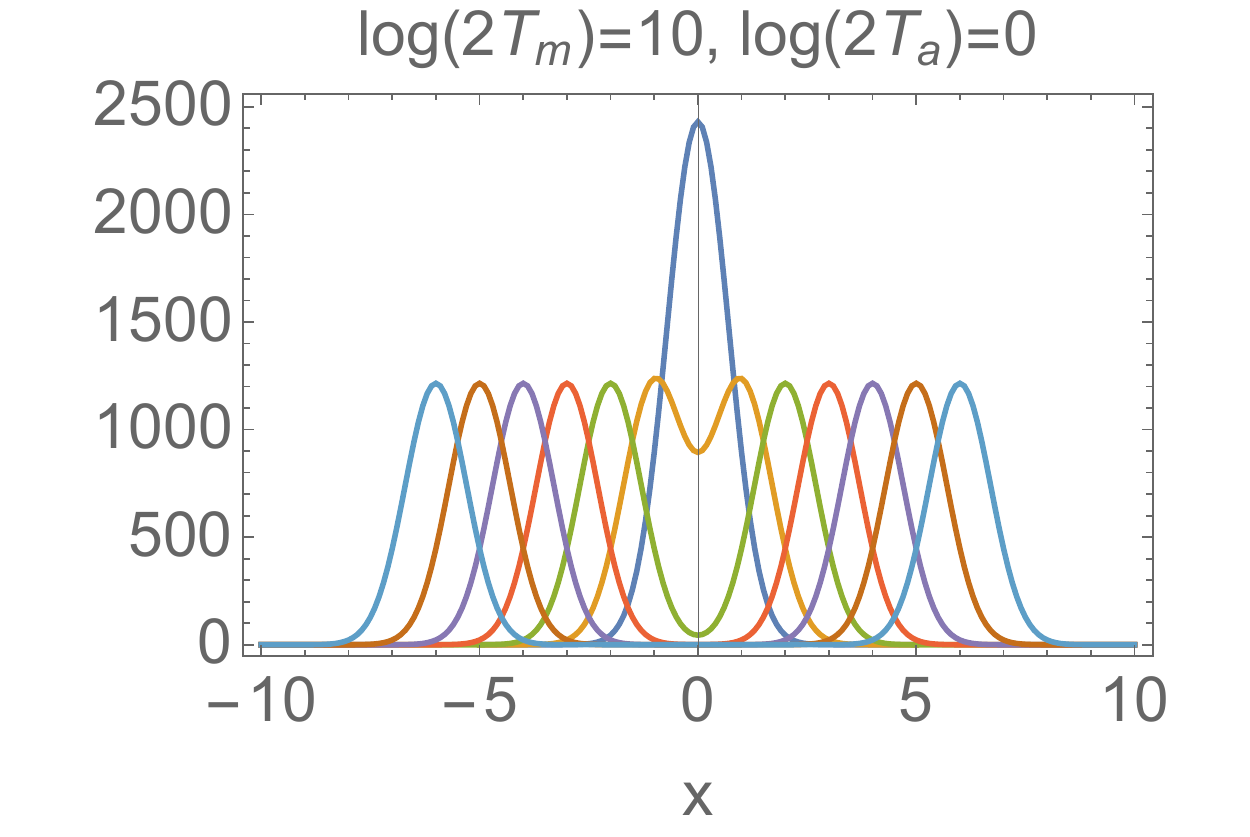}
    \caption{The function $\texttt{q}_0^{\rm FF}(x,t)$ as a function of $x$ for small values of $t=0,0.2,0.4,0.6,0.8$ and $1$  (Left) and for larger values $t=0,1,2,3,4,5$ and $6$ (Right). }
    \label{freefig}
\end{figure}

 It is well-known that in a free fermion, the time evolution of $\texttt{q}_0^{\rm FF}(x,t)$ is characterised by a double-bump release, whereby for $t=0$ the particle density is localised around $x=0$ and then splits into two identical peaks which propagate asymptotically at speeds $\pm 1$ producing the curves in Fig.~\ref{freefig}.
As we can see in Fig.~\ref{freefig} (left) the initial local maximum of the density at $x=0$ first reduces and then splits into two. For early times, the value of $\texttt{q}_0^{\rm FF}(0,t)$ can be analytically obtained from (\ref{FFq0}) as
\beq
\texttt{q}_0^{\rm FF}(0,t)=\frac{1}{2\pi}\int_{-\infty}^\infty \frac{\cosh\theta}{1+e^{\beta_m e^{v(\theta)^2 t^2}\cosh\theta}}\,d\theta=  \frac{1}{2\pi} \sum_{n=0}^\infty \int_{-\infty}^\infty {\cosh\theta} (-1)^n e^{-(n+1) \beta_m e^{v(\theta)^2 t^2}\cosh\theta}\,,
\eeq
where $v(\theta)=\tanh\theta$ and $\beta_m=1/T_m$ and for simplicity we take $T_a=0$. Then, approximating $v(\theta)\approx \pm 1$ and expanding, the above equals
\beq
\frac{1}{\pi} \sum_{n=0}^\infty \int_{0}^\infty {\cosh\theta} (-1)^n e^{-(n+1) \beta_m e^{t^2}\cosh\theta}= \frac{1}{\pi} \sum_{n=0}^\infty (-1)^n K_1((n+1) \beta_m e^{t^2})\,,
\eeq
and for high temperatures this can be approximated by
\beq
\texttt{q}_0^{\rm FF}(0,t)\approx \frac{1}{\pi \beta_m e^{t^2}} \sum_{n=0}^\infty  
\frac{(-1)^n}{n+1}= \frac{\log 2}{\pi \beta_m e^{t^2}}=q^{\rm FF}_0(0,0) e^{-t^2}.
\label{1r}
\eeq
In conclusion, for small times the maximum of the free fermion density bump reduces as a gaussian. A similar analysis gives
\beq
\texttt{q}_0^{\rm FF}(x,0)\approx \texttt{q}_0^{\rm FF}(0,0) e^{-x^2}.
\label{2r}
\eeq
 As we will see in Appendix~\ref{App:B}, for large temperature $T_m$, this result generalises to interacting theories where $\log 2$ is replaced by the value $L(0)$ of the TBA $L$-function. It is easy to show by similar methods, that the height of the two peaks that are produced after the release of the initial density bump in Fig.~\ref{freefig} is exactly half that of the original bump, that is
 \beq 
 \lim_{t \rightarrow \infty}\texttt{q}^{\rm FF}_0(\pm t,t)= \frac{\texttt{q}^{\rm FF}_0(0,0)}{2}\,.
 \eeq 
 From this result it can be estimated that the bump splitting occurs roughly for a time $t_g$ such that $e^{-t_g^2}=\frac{1}{2}$, that is for $t_g=\sqrt{\log2}=0.8325...$. This value is interesting because it represents the end of the region which we have referred to as ``early times" and the start of the region we have called  ``intermediate times" throughout the paper. This value is of course only computed here for a free fermion and will be different for an interacting theory, but it gives a good indication of the order of magnitude. 
 
 Finally, it is a simple matter to show that when $T_a\neq 0$ the result (\ref{1r}) generalises by replacing
 \beq
 e^{-t^2} \mapsto R+(1-R)e^{-t^2}\qquad \mathrm{for} \qquad  R=\frac{T_a}{T_m}\,.
 \eeq
The generalization of these results to interacting theories provides an analytical justification for our early time results, where gaussian decay dominates the early-time decay of the subsidiary peak both in the presence and absence of a bath. 

\section{Computing ${\tt q}^+_0(0,0)$ Analytically for Large  Temperatures}
\label{App:B}
 Let us look at the particle density  at equilibrium at temperature $T$ for a simple interacting theory, consisting of a single particle type and a diagonal $S$-matrix. By definition
\beq
       \text{q}_0(0,0) =\frac{1}{2\pi}\int_{-\infty}^\infty  n(\theta) e^{\rm dr}(\theta) \, d\theta.
\eeq
and we define the useful variable
\beq
y:=\log\frac{2}{\beta}=\log(2T)\,.
\eeq
Our aim is to study the asymptotics of this value as a function of temperature for $y\rightarrow \infty$, that is $T\gg 1$. Although this is a very standard TBA computation, we believe it has not been presented elsewhere, although it would follow straighforwardly from the formulae for generic spin densities presented in \cite{ourroaming} (for instance).
Let us rewrite this integral as 
    \beqa
       \text{q}_0(0,0) &=& \frac{1}{2\pi}\int_{-\infty}^0  n(\theta) e^{\rm dr}(\theta) \, d\theta+ \frac{1}{2\pi}\int_{0}^\infty  n(\theta) e^{\rm dr}(\theta) \, d\theta \nonumber\\
      & = &    \frac{1}{2\pi}\int_{-\infty}^{y}  n(\theta-y) e^{\rm dr}(\theta-y) \, d\theta + \frac{1}{2\pi}\int_{-y}^{\infty}  n(\theta+y) e^{\rm dr}(\theta+y) \, d\theta 
    \eeqa
For a generic theory with a single particle, we can compute $\varepsilon(\theta \pm y):=\varepsilon_{\pm}(\theta)$ in the limit of $y$ large by writing the TBA equation 
(note that the $\pm$ indices here have nothing to do with the HSG model)
\beq
\varepsilon(\theta)=\beta \cosh(\theta) -( \varphi \star L)(\theta)= e^{\theta-y}+e^{-\theta-y}-( \varphi \star L)(\theta)\,,
\eeq
so that
\beq
\varepsilon_{\pm}(\theta)=e^{\theta-y\pm y}+e^{-\theta-y\mp y}-( \varphi \star L)(\theta\pm x)\approx e^{\pm \theta} - (\varphi \star L_{\pm})(\theta)\,,\quad y\gg 1\,,
\eeq
and we also have that, differentiating this equation,
\beq
\varepsilon_{\pm}'(\theta)=\pm e^{\pm \theta}-(\varphi'\star L_\pm)(\theta) \approx \pm e^{\pm \theta}-(\varphi \star L_\pm')(\theta)\,,
\label{here}
\eeq
where the last step follows from integration by parts and using the fact that $\varphi(\theta)$ should vanish at $\pm \infty$ and the identity $\frac{d}{d\theta}\varphi(\theta-\theta')=-\frac{d}{d\theta'}\varphi(\theta-\theta')$. By definition
\beq
L_\pm'(\theta)=-n_\pm(\theta) \varepsilon'_\pm(\theta)\,,
\eeq
and, if we call $\varepsilon'_\pm(\theta):=[\pm e^{\pm \theta}]^{\rm dr}$ then the equation above becomes exactly the equation for the dressing of the function $\pm e^{\pm \theta}$. On the other hand, the equation for the dressing of the energy is very similar to (\ref{here})
\beq
e^{\rm dr}(\theta)=\cosh\theta+(\varphi \star n e^{\rm dr})(\theta)\,,
\label{here2}
\eeq
and if we shift as above we have that, for $y\gg 1$
\beq
e^{\rm dr}_\pm(\theta)\approx \frac{e^{\pm \theta+y}}{2}+(\varphi \star n_\pm e_\pm^{\rm dr})(\theta)\,.
\label{here3}
\eeq
Comparing (\ref{here}) and (\ref{here2}) we have that they are essentially identical if we identify $  \varepsilon_\pm'(\theta)=\pm \beta e^{\rm dr}_\pm (\theta)$. This allows us to do the following replacement in the particle density:
\beqa
       \text{q}_0(0,0) & = &    \frac{1}{2\pi}\int_{-\infty}^{y}  n(\theta-y) e^{\rm dr}(\theta-y) \, d\theta + \frac{1}{2\pi}\int_{-y}^{\infty}  n(\theta+y) e^{\rm dr}(\theta+y) \, d\theta \nonumber\\
       &\approx &   \frac{e^y}{4\pi}\int_{-\infty}^{y}  n(\theta-y) [-e^{-\theta}]^{\rm dr}\ \, d\theta + \frac{e^y}{4\pi}\int_{-y}^{\infty}  n(\theta+y) [e^{\theta}]^{\rm dr}\ \, d\theta \nonumber\\
        &\approx &   -\frac{e^y}{4\pi}\int_{-\infty}^{y}  n(\theta-y) \varepsilon'(\theta-y)\ \, d\theta + \frac{e^y}{4\pi}\int_{-y}^{\infty}  n(\theta+y) \varepsilon'(\theta+y) \, d\theta \nonumber\\
         &\approx &   \frac{1}{2\beta \pi}\int_{-\infty}^{y}  L'(\theta-y)\ \, d\theta - \frac{1}{2\beta \pi}\int_{-y}^{\infty}  L'(\theta+y) \, d\theta= \nonumber\\
        &=&\frac{1}{2\beta \pi} (L(0)-L(-\infty-y)-L(\infty+y)+L(0)).
    \eeqa
    For $y \rightarrow \infty$ we have that $L(\pm \infty\pm y)\rightarrow 0$ (the $L$-functions typically go to zero at $\pm \infty$ and have a plateau between $-y$ and $y$) so the final result is
    \beq
        \text{q}_0(0,0) \approx  \frac{L(0)}{\beta \pi}\,.
        \label{res47}
    \eeq
     $L(0)$ for $y$ large is the middle of the plateau of the $L$-function, so it will be model-dependent. It is easy to shown that $L(0)=\log 2$ for the free fermion.
    The result (\ref{res47}) generalises easily to the case when $\beta$ is a function of $x$, giving (\ref{smallt}). 
   
   \subsection{Spectral Density Local Maxima at Equilibrium}\label{bath_height}
A related but even simpler computation allows us to find approximate formulae for the height of the peaks of the spectral density function $\rho^{\text{FF}}_p(\theta)$ at equilibrium. It is very easy to show that, for $T\gg 1$, the function 
\beq
\rho^{\text{FF}}_p(\theta)=\frac{1}{2\pi} \frac{\cosh\theta}{1+e^{\beta \cosh\theta}}\,,
\eeq 
has local maxima at $\theta=\pm \log(2T)$ with $\beta=1/T$. The value of the function at each of these maxima can be obtained by substituting $\theta=\pm\log(2T)$. Then, expanding around $T$ large (or $\beta$ small) we can obtain the leading dependence in $T$. Such expansion gives
\beq
h:=\rho^{\text{FF}}_p(\pm \log(2T))=\frac{T}{2\pi(1+e)}+\frac{1}{4\pi T(1+e)^2}+\mathcal{O}(T^{-2})\,.
\eeq
This means that if we call $h$ the height of the maximum of either of these peaks, we have that:
\beq
\log h \approx \log(2T)-\log(4\pi(1+e))= \log(2T)-3.8443\,.
\label{fitridge}
\eeq
This formula is recovered numerically in our Fig.\ref{hs_Vs}, top right panel where the fit of the height of the bath ridges is quite precisely given by (\ref{fitridge}).
    \section{Tables of Fitting Parameters}
    \label{fitt}



\begin{table}[h!]              
\centering                                                   {\small
\begin{tabular}{|c||c|c|c||c|c|}
\hline                                                       Function  & Fits & Parameters & $\Delta$ & Time Interval &$R^2$ \\   
\hline\hline 
 $\Gamma(t)$ & $f_1(t)=c_\Gamma\, t+d_\Gamma$ &  $c_\Gamma=-0.001$ & $\Delta c_\Gamma=  0.01$ & [1.00 , 4.75] & 0.0015 \\  
&  &   $d_\Gamma=0.10$ & $\Delta d_\Gamma=   0.02$ &  &\\    
\hline
 $\Gamma(t)$ &  $f_2(t)=\frac{2(t-b_\Gamma)}{a_\Gamma}$  & $a_\Gamma=32.36$ &  $\Delta a_\Gamma=1.56$ & [5.00 , 10.00] &  0.9900 \\
&  &  $b_\Gamma=3.89$ & $\Delta b_\Gamma=0.19$ & & \\ 
\hline
 $\log \rho^+_p (t)$  & $g_1(t)=-\frac{t-p_\rho}{q_\rho}$ & $p_\rho=22.50$ & $\Delta p_\rho=0.49$ & [2.00 , 5.00] & 0.9985 \\
  &  &   $q_\rho= 11.53$ & $\Delta q_\rho=  0.30$ & &    \\
\hline      
 $\log \rho^+_p (t)$   & $g_2(t)-\frac{(t-b_\rho)^2}{a_\rho} + h_\rho$  & $b_\rho=4.19$ & $\Delta b_\rho=0.12$ &[4.75 , 10.00] & 0.9997 \\  
  &  &   $a_\rho= 33.31$ & $\Delta a_\rho= 1.18$ & &  \\
 &  &   $h_\rho=1.54$ & $\Delta h_\rho=0.01$ &  & \\   
\hline                                                  
\end{tabular}}                                                             
\caption{Fitting parameters for the functions $\Gamma(t)$ and $\log \rho^+_p (t)$ shown in Fig.~\ref{fig3}. As expected from equation (\ref{keye}) we have that $b_\rho \approx b_\Gamma$, $a_\rho \approx a_\Gamma$ and $q_\rho \approx 1/d_\Gamma \approx 10$.}
\label{table:fit_decay}                                                 
\end{table}

 \begin{table}[h!]
\centering       
{\small 

\begin{tabular}{|c|c||c|c|c|c|c|c|c|}     
\cline{1-7}
$\mathbf{\sigma}$ & $\mathbf{\alpha_m}$ & $\tau_H$ & $\Delta \tau_H$  &  $\xi_H$ & $\Delta \xi_H$  &$R^2$ \\
\hline  
\hline                                            
12.00 & 7.50 & 1.72 & 0.05 & 4.06 & 0.05 & 0.9989 \\
\hline                                            
14.00 & 8.50 & 1.71 & 0.05 & 5.07 & 0.05 & 0.9988 \\
\hline                                            
16.00 & 9.50 & 1.71 & 0.05 & 6.07 & 0.05 & 0.9988 \\
\hline                                            
\end{tabular}  }

\caption{Linear fit parameters $\log H_0^+(t)= -\frac{t}{\tau_H} + \xi_H$  for $t \in [2, 4]$ for $\kappa=1.5$ in zero bath.}                  \label{tab:kappa15}  
 \end{table}

\begin{table}[h!]
\centering       
{\small 

\begin{tabular}{|c|c||c|c|c|c|c|c|c|}     
\cline{1-7}
$\mathbf{\sigma}$ & $\mathbf{\alpha_m}$ & $\tau_Q$ & $\Delta \tau_Q$  &  $\xi_Q$ & $\Delta \xi_Q$  &$R^2$ \\
\hline  
12.00 & 7.50 & 2.75 & 0.02 & 4.20 & 0.01  & 0.9997 \\
\hline                                                          
14.00 & 8.50 & 2.65 & 0.03 & 5.24 & 0.02  & 0.9995 \\
\hline                                                          
16.00 & 9.50 & 2.62 & 0.03 & 6.26 & 0.02  & 0.9993 \\
 
\hline  
\end{tabular}}
\caption{Linear fit parameters $\log Q_0^+(t)=-\frac{t}{\tau_Q} + \xi_Q$ for  $t \in [2, 7.5]$ for $\kappa=1.5$ in zero bath. }     
\label{tab:kappa15Q}  
 \end{table}


\begin{table}[h!]
\begin{center}
\begin{tabular}{|c|c|c|c||c|c||c|c||c|c|}    

\hline
$\kappa$ & $\sigma$ & $\mathbf{\alpha_m}$ & $\Delta \theta_{max}$ & $\tau_H$ & $\Delta \tau_H$ & $\xi_H$ & $\Delta \xi_H$  & time int. & $R^2$ \\

\hline  
\hline
1.40 & 14.00 & 8.40 & 0.08 & 1.43 & 0.05 & 0.07 & 0.07  & [2.00,3.50] & 0.9990 \\  
\hline                                                                                       
1.50 & 14.00 & 8.50 & 0.08 & 1.71 & 0.05 & -0.09 & 0.05 & [2.00,4.00] & 0.9988 \\ 
\hline                                                                                       
1.60 & 14.00 & 8.60 & 0.08 & 2.13 & 0.03 & -0.26 & 0.02 &[2.00,4.00] & 0.9998 \\ 
\hline                                                                                       
1.75 & 14.00 & 8.75 & 0.09 & 2.82 & 0.03 & -0.42 & 0.01  &[2.25,5.25] & 0.9998 \\ 
\hline                                                                                       
1.90 & 14.00 & 8.90 & 0.08 & 3.67 & 0.06 & -0.53 & 0.02 & [2.25,7.25] &0.9987 \\ 
\hline                                                                                       
2.00 & 16.00 & 10.00 & 0.12 & 4.53 & 0.06 & -0.61 & 0.02  & [2.25,7.25] & 0.9992 \\
\hline                                                                                       
2.10 & 14.00 & 9.10 & 0.08 & 5.60 & 0.05 & -0.69 & 0.01 & [3.25,8.00] & 0.9997 \\ 
\hline                                                                                       
2.25 & 14.00 & 9.25 & 0.08 & 7.29 & 0.07 & -0.73 & 0.01 & [2.25,12.50] & 0.9992 \\ 
\hline                                                                                       
2.50 & 14.00 & 9.50 & 0.09 & 12.12 & 0.06 & -0.83 & 0.01  & [3.50, 20.00] & 0.9996 \\
\hline                                                                                    
\end{tabular}     
\end{center}
\caption{Parameters of the fitting functions $-\frac{t}{\tau_H} +\xi_H$ for various $\kappa$ and $T_a =0$. The fitting functions for $\kappa = 1.50, 1.75, 2.00$ and $2.25$ are 
shown in Fig.~\ref{fig4}. $\Delta \theta_{max}$ indicates the best rapidity resolution, and  $\tau_H \pm \Delta \tau_H $ and $\xi_H \pm \Delta \xi_H $ correspond to the $95\%$-confidence bounds of the two coefficients. The coefficients $\tau_H$ are fitted in figure Fig.~\ref{fig5} with fit parameters shown in Table~ \ref{table4}.} 
\label{manyka}
\end{table}

\begin{table}[h!]
\begin{center}
\begin{tabular}{|c|c|c|c||c|c||c|c||c|c|}    

\hline
$\kappa$ & $\sigma$ & $\mathbf{\alpha_m}$ & $\Delta \theta_{max}$ & $\tau_Q$ & $\Delta \tau_Q$ & $\xi_Q$ & $\Delta \xi_Q$  & time int. & $R^2$ \\
\hline\hline
1.00 & 14.00 & 8.00 & 0.08 & 1.07 & 0.07 & 1.28 & 0.19 & [1.75,4.50] & 0.9923 \\  
\hline                                                                                       
1.25 & 14.00 & 8.25 & 0.09 & 1.72 & 0.04 & 0.71 & 0.05 & [2.00,5.25] & 0.9989 \\  
\hline                                                                                       
1.40 & 14.00 & 8.40 & 0.08 & 2.22 & 0.03 & 0.51 & 0.03 & [2.00,6.50] &  0.9994 \\  
\hline                                                                                       
1.50 & 14.00 & 8.50 & 0.08 & 2.65 & 0.03 & 0.39 & 0.02 & [2.00,7.50] & 0.9995 \\  
\hline                                                                                       
1.60 & 14.00 & 8.60 & 0.08 & 3.18 & 0.03 & 0.28 & 0.02  & [2.25,8.75] & 0.9996 \\  
\hline                                                                                       
1.75 & 14.00 & 8.75 & 0.09 & 4.23 & 0.03 & 0.13 & 0.01 & [2.75,11.00] & 0.9996 \\  
\hline                                                                                       
1.90 & 14.00 & 8.90 & 0.08 & 5.63 & 0.03 & 0.01 & 0.01  & [3.25,14.25] & 0.9996 \\  
\hline                                                                                       
2.00 & 16.00 & 10.00 & 0.12 & 6.74 & 0.04 & -0.06 & 0.01 & [3.75,16.50] & 0.9996 \\
\hline                                                                                       
2.10 & 14.00 & 9.10 & 0.08 & 8.45 & 0.06 & -0.15 & 0.01 & [5.25,15.00] &  0.9996 \\ 
\hline                                                                                       
2.25 & 14.00 & 9.25 & 0.08 & 11.33 & 0.07 & -0.23 & 0.01& [6.25,20.00] &  0.9995 \\
\hline                                                                                       
2.50 & 14.00 & 9.50 & 0.09 & 17.52 & 0.17 & -0.28 & 0.01  & [8.75,20.00] & 0.9990 \\
\hline                                                                                            
\end{tabular}
\end{center}
\caption{Parameters of the fitting functions $-\frac{t}{\tau_Q} +\xi_Q$ for various $\kappa$ and $T_a =0$. The fitting functions for $\kappa =1, 1.25, 1.50, 1.75, 2$ and $2.25$ are shown in Fig.~\ref{fig4}. $\Delta \theta_{max}$ indicates the best rapidity resolution, and  $\tau_Q \pm \Delta \tau_Q $ and $\xi_Q \pm \Delta \xi_Q $ correspond to the $95\%$-confidence bounds of the two coefficients. The coefficients $\tau_Q$ are fitted as displayed in figure Fig.~\ref{fig5} with fit parameters shown in in Table~\ref{table4}. 
}
\label{manykaQ}
\end{table}

\begin{table}[h!]                                                  
\centering                                                     
\begin{tabular}{|c||c|c|c|c|c|c|}                               
\hline   
Function & $c_{H,Q}$ & $ \Delta c_{H,Q}$ & $d_{H,Q}$ & $ \Delta d_{H,Q}$  & $R^2$ \\
\hline \hline 
$\log \tau_H (\kappa) $  & 1.94 & 0.03 & 1.22 & 0.01 &  0.9997 \\
\hline
$\log \tau_Q (\kappa) $  & 1.88 & 0.03 & 0.97 & 0.01 & 0.9995 \\
\hline                                                         
\end{tabular}                                                  
\caption{Fits of $\log \tau_H (\kappa)$ and $\log \tau_Q (\kappa)$. The fitting functions $ c_H( \kappa - d_H)$ and $ c_Q( \kappa - d_Q)$ are shown in  Fig.~\ref{fig5}, and the fitted values are displayed in Table~\ref{manyka} and in Table~\ref{manykaQ} respectively. $\Delta c_{H,Q} $ and $\Delta d_{H,Q} $ denote the $95\%$-confidence intervals of the two coefficients.   
}                                  
\label{table4}                                      
\end{table}


\begin{table}[h!]
\centering       
{\small                                                                 
\begin{tabular}{|c||c|c|c|c|c|}     
\cline{1-6}
${\sigma}$ & $a'_m$ & $\Delta a'_m$  &  $b'_m$ & $\Delta b'_m$  & $R^2$ \\
\hline  
\hline           
12 & 0.084 & 0.004 &  -3.65 & 0.24 & 0.9979\\   
\hline 

10 & 0.078 & 0.005 &  -2.61 & 0.22 & 0.9970 \\   
\hline   
\end{tabular}}   

\caption{Parameter of the fit (\ref{hsfit}) for two values of the resonant parameter $\sigma=12,10$ and  bath temperature exponent $\alpha_a=3$. $\Delta a'_m, \Delta b'_m$ are the uncertainties of parameters $a'_m,b'_m$ respectively. $R^2$ is the R-squared goodness of fit parameter.}                  \label{tab:h1}  
 \end{table}
 


 \begin{table}[h!]
\centering       
{\small                                                                 
\begin{tabular}{|c||c|c|c|c|c|c|c|c|c|}     
\cline{1-8}
$\mathbf{\sigma}$ & $a'_a$ & $\Delta a'_a$ & $b'_a$ & $\Delta b'_a$ & $c'_a$ & $\Delta c'_a$  &$R^2$ \\
\hline  
\hline           
12 & -0.12 & 0.01 & 1.41 & 0.05 & -2.14 & 0.05 &  0.9995\\   
\hline 
 
\end{tabular}}   

\caption{Parameter of the fit (\ref{hsqua}) for resonance parameter $\sigma=12$ and maximum temperature exponent $\alpha_m=7.5$. $\Delta a'_b$ is the uncertainty of the parameter $a'_b$. $R^2$ is the R-squared goodness of the fit.}                  \label{tab:h2}  
 \end{table}
 

\begin{table}[h!]
\centering       
{\small                                                                 
\begin{tabular}{|c||c|c|c|}     
\cline{1-4}
${\sigma}$ & $a'_b$ & $\Delta a'_b$  & $R^2$ \\
\hline  
\hline           
12 & -3.80 & 0.03  & 0.9991\\   
\hline 

\end{tabular}}   

\caption{Parameters of the fit (\ref{hbfit}) for $\sigma=12$ and  $\alpha_m=7.5$. $\Delta a'_b$ is the uncertainty of parameter $a'_b$. $R^2$ is the R-squared goodness of the fit.}                  \label{tab:hb}  
 \end{table}

 \begin{table}[h!]
\centering       
{\small                                                                 
\begin{tabular}{|c||c|c|c|c|c|c|c|}     
\cline{1-6}
$\mathbf{\sigma}$ & $a''_m$ & $\Delta a''_m$  &  $b''_m$ & $\Delta b''_m$  &$R^2$ \\
\hline  
\hline           
12 & 1.72 & 0.08 &  -10.87 & 0.66 & 0.9971\\   
\hline 

10 & 1.58 & 0.09 &  -8.52 & 0.58 & 0.9977 \\   
\hline

\end{tabular}}   

\caption{Parameters of the fit (\ref{vollin}) with respect to $\alpha_m$ for two values of the resonance parameter $\sigma=12,10$ and bath temperature exponent $\alpha_a=3$. $\Delta a''_m, \Delta b''_m$ are the uncertainties of parameters $a''_m,b''_m$ respectively. $R^2$ is the R-squared goodness of the fit.}                  \label{tab:V1}  
 \end{table}
 \begin{table}[h!]
\centering       
{\small                                                                 
\begin{tabular}{|c||c|c|c|c|c|c|c|c|c|}     
\cline{1-8}
$\mathbf{\sigma}$ & $a''_a$ & $\Delta a''_a$  &  $b''_a$ & $\Delta b''_a$ &  $c''_a$ & $\Delta c''_a$  &$R^2$ \\
\hline  
\hline           
12 & -0.075 & 0.008 &  1.26 & 0.04 & -0.98 & 0.04  & 0.9998\\   
\hline 
 
\end{tabular}}   

\caption{Parameters of the fit (\ref{vollin}) with respect to $\alpha_a$ for resonance parameter $\sigma=12$ and maximum temperature exponent $\alpha_m=7.5$. $\Delta a''_a, \Delta b''_a, \Delta c''_a$ are the uncertainties of parameters $a''_a,b''_a,c''_a$ respectively. $R^2$ is the R-squared goodness of the fit.}                  \label{tab:V2}  
 \end{table}


\newpage
\begin{thebibliography}{10}

\bibitem{Breit-Wigner}
G.~Breit and E.~Wigner,
\newblock Capture of Slow Neutrons,
\newblock Phys. Rev. {\bf 49}, 519--531 (Apr 1936).

\bibitem{WZNW1}
J.~Wess and B.~Zumino,
\newblock {Consequences of anomalous Ward identities},
\newblock Phys. Lett.  {\bf B37}, 95--97 (1971).

\bibitem{WZNW2}
S.~Novikov,
\newblock {Multivalued functions and functionals. An analogue of the Morse
  theory},
\newblock Sov. Math. Dokl. {\bf 24}, 222 (1981).

\bibitem{WZNW3}
S.~Novikov,
\newblock {The Hamiltonian formalism and a many-valued analogue of Morse
  theory},
\newblock Russ. Math. Sur. {\bf 37}, 1 (1982).

\bibitem{WZNW4}
E.~Witten,
\newblock {Global Aspects of Current Algebra},
\newblock Nucl. Phys.  {\bf B223}, 422--432 (1983).

\bibitem{WZNW5}
E.~Witten,
\newblock {Non-abelian bosonization in two dimensions},
\newblock Comm. Math. Phys {\bf 92}, 455--472 (1984).

\bibitem{hsg}
C.~R. Fernandez-Pousa, M.~V. Gallas, T.~J. Hollowood, and J.~L. Miramontes,
\newblock {Solitonic integrable perturbations of parafermionic theories},
\newblock Nucl. Phys. {\bf B499}, 673--689 (1997).

\bibitem{ntft}
C.~Fernandez-Pousa, M.~Gallas, T.~Hollowood, and J.~Miramontes,
\newblock {The symmetric space and homogeneous sine-Gordon theories},
\newblock Nucl. Phys. {\bf B484}, 609--630 (1997).

\bibitem{FernandezPousa:1997iu}
C.~R. Fernandez-Pousa and J.~L. Miramontes,
\newblock {Semi-classical spectrum of the homogeneous sine-Gordon theories},
\newblock Nucl. Phys. {\bf B518}, 745--769 (1998).

\bibitem{smatrix}
J.~L. Miramontes and C.~R. Fernandez-Pousa,
\newblock {Integrable quantum field theories with unstable particles},
\newblock Phys. Lett. {\bf B472}, 392--401 (2000).

\bibitem{tba1}
A.~Zamolodchikov,
\newblock {Thermodynamic Bethe ansatz in relativistic models. Scaling three
  state Potts and Lee-Yang models},
\newblock Nucl. Phys. {\bf B342}, 695--720 (1990).

\bibitem{tba2}
T.~R. Klassen and E.~Melzer,
\newblock {The Thermodynamics of purely elastic scattering theories and
  conformal perturbation theory},
\newblock Nucl. Phys. {\bf B350}, 635--689 (1991).

\bibitem{KW}
M.~Karowski and P.~Weisz,
\newblock Exact S matrices and form-factors in (1+1)-dimensional field
  theoretic models with soliton behavior,
\newblock Nucl. Phys. {\bf B139}, 455--476 (1978).

\bibitem{SmirnovBook}
F.~Smirnov,
\newblock Form factors in completely integrable models of quantum field theory,
\newblock Adv. Series in Math. Phys. {\bf 14}, World Scientific, Singapore (1992).

\bibitem{ourtba}
O.~A. Castro-Alvaredo, A.~Fring, C.~Korff, and J.~L. Miramontes,
\newblock {Thermodynamic Bethe ansatz of the homogeneous sine-Gordon models},
\newblock Nucl. Phys. {\bf B575}, 535--560 (2000).

\bibitem{CastroAlvaredo:2002nv}
O.~A. Castro-Alvaredo, J.~Dreissig, and A.~Fring,
\newblock {Integrable scattering theories with unstable particles},
\newblock Eur. Phys. J. {\bf C35}, 393--411 (2004).

\bibitem{Dorey:2004qc}
P.~E. Dorey and J.~L. Miramontes,
\newblock {Mass scales and crossover phenomena in the homogeneous sine-Gordon
  models},
\newblock Nucl. Phys. {\bf B697}, 405--461 (2004).

\bibitem{CastroAlvaredo:2000em}
O.~A. Castro-Alvaredo, A.~Fring, and C.~Korff,
\newblock {Form factors of the homogeneous sine-Gordon models},
\newblock Phys. Lett. {\bf B484}, 167--176 (2000).

\bibitem{CastroAlvaredo:2000nk}
O.~A. Castro-Alvaredo and A.~Fring,
\newblock {Identifying the operator content, the homogeneous sine- Gordon
  models},
\newblock Nucl. Phys. {\bf B604}, 367--390 (2001).

\bibitem{CastroAlvaredo:2000ag}
O.~A. Castro-Alvaredo and A.~Fring,
\newblock {Renormalization group flow with unstable particles},
\newblock Phys. Rev. {\bf D63}, 021701 (2001).

\bibitem{CastroAlvaredo:2000nr}
O.~A. Castro-Alvaredo and A.~Fring,
\newblock {Decoupling the $SU(N)_2$ homogeneous sine-Gordon model},
\newblock Phys. Rev. {\bf D64}, 085007 (2001).

\bibitem{Bajnok1}
Z.~Bajnok, J.~Balog, K.~Ito, Y.~Satoh, and G.~Z. Toth,
\newblock {On the mass-coupling relation of multi-scale quantum integrable
  models},
\newblock JHEP {\bf 06}, 071 (2016).

\bibitem{Bajnok2}
Z.~Bajnok, J.~Balog, K.~Ito, Y.~Satoh, and G.~Z. T\'oth,
\newblock {Exact mass-coupling relation for the homogeneous sine-Gordon model},
\newblock Phys. Rev. Lett. {\bf 116}(18), 181601 (2016).

\bibitem{ourUP}
O.~A. Castro-Alvaredo, C.~De~Fazio, B.~Doyon, and F.~Ravanini,
\newblock {On the hydrodynamics of unstable excitations},
\newblock JHEP {\bf 09}, 045 (2020).

\bibitem{tails}
O.~A. Castro-Alvaredo, C.~De~Fazio, B.~Doyon, and A.~A. Zi\'o\l{}kowska,
\newblock Tails of Instability and Decay: a Hydrodynamic Perspective,
\newblock arXiv: 2103.03735  (2021).

\bibitem{ourhydro}
O.~A. Castro-Alvaredo, B.~Doyon, and T.~Yoshimura,
\newblock {Emergent hydrodynamics in integrable quantum systems out of equilibrium},
\newblock Phys. Rev. {\bf X6}(4), 041065 (2016).

\bibitem{theirhydro}
B.~Bertini, M.~Collura, J.~De~Nardis, and M.~Fagotti,
\newblock {Transport in Out-of-Equilibrium $XXZ$ Chains: Exact Profiles of
  Charges and Currents},
\newblock Phys. Rev. Lett. {\bf 117}(20), 207201 (2016).

\bibitem{Brev}
B.~Doyon,
\newblock Lecture notes on Generalised Hydrodynamics,
\newblock SciPost Phys. Lecture Notes  (2020).

\bibitem{20}
R.~Zwanzig,
\newblock {Non-equilibrium Statistical Physics},
\newblock New York, Oxford University Press  (2001).

\bibitem{Eisert}
K.~Audenaert, J.~Eisert, M.~B. Plenio, and R.~F. Werner,
\newblock Entanglement properties of the harmonic chain,
\newblock Phys. Rev. {\bf A66}, 042327 (2002).

\bibitem{NERev2}
F.~H.~L. Essler and M.~Fagotti,
\newblock Quench dynamics and relaxation in isolated integrable quantum spin
  chains,
\newblock J. Stat. Mech. {\bf 2016}(6), 064002 (2016).

\bibitem{NERev3}
R.~Vasseur and J.~E. Moore,
\newblock Nonequilibrium quantum dynamics and transport: from integrability to
  many-body localization,
\newblock J. Stat. Mech. {\bf 2016}(6), 064010 (2016).

\bibitem{NERev4}
E.~Ilievski, M.~Medenjak, T.~Prosen, and L.~Zadnik,
\newblock Quasilocal charges in integrable lattice systems,
\newblock J. Stat. Mech. {\bf 2016}(6), 064008 (2016).

\bibitem{NERev5}
D.~Bernard and B.~Doyon,
\newblock Conformal field theory out of equilibrium: a review,
\newblock J. Stat. Mech. {\bf 2016}(6), 064005 (2016).

\bibitem{NERev6}
L.~D'Alessio, Y.~Kafri, A.~Polkovnikov, and M.~Rigol,
\newblock From quantum chaos and eigenstate thermalization to statistical
  mechanics and thermodynamics,
\newblock Adv. in Phys. {\bf 65}(3), 239--362 (2016).

\bibitem{NERev7}
C.~Gogolin and J.~Eisert,
\newblock Equilibration, thermalisation, and the emergence of statistical
  mechanics in closed quantum systems,
\newblock Reports on Progress in Physics {\bf 79}(5), 056001 (2016).

\bibitem{CEM}
P.~Calabrese, H.~Essler, and G.~Mussardo~(ed.),
\newblock {Quantum Integrability in Out-of-Equilibrium Systems},
\newblock J. Stat. Phys. 064001 (2016).

\bibitem{kinoshita}
T.~Kinoshita, T.~Wenger, and D.~Weiss,
\newblock {A Quantum Newton's Cradle},
\newblock Nature {\bf 440}, 900 (2006).

\bibitem{Rigol}
M.~Rigol, V.~Dunjko, V.~Yurovsky, and M.~Olshanii,
\newblock Relaxation in a Completely Integrable Many-Body Quantum System: An Ab
  Initio Study of the Dynamics of the Highly Excited States of 1D Lattice
  Hard-Core Bosons,
\newblock Phys. Rev. Lett. {\bf 98}, 050405 (2007).

\bibitem{iFluid}
F.~S. M\o{}ller and J.~Schmiedmayer,
\newblock Introducing iFluid: a numerical framework for solving hydrodynamical
  equations in integrable models,
\newblock SciPost Physics {\bf 8}(3) (Mar 2020).

\bibitem{iFluid2}
F.~S. M\o{}ller, G.~Perfetto, B.~Doyon, and J.~Schmiedmayer,
\newblock Euler-scale dynamical correlations in integrable systems with fluid
  motion,
\newblock SciPost Physics Core {\bf 3}(2) (Dec 2020).

\bibitem{FM}
D.~Fioretto and G.~Mussardo,
\newblock Quantum quenches in integrable field theories,
\newblock New J. Phys. {\bf 12}(5), 055015 (2010).

\bibitem{Mossel}
J.~Mossel and J.-S. Caux,
\newblock Generalized TBA and generalized Gibbs,
\newblock J. Phys. {\bf A45}, 255001 (2012).

\bibitem{BEL14}
L.~Bonnes, F.~H.~L. Essler, and A.~M. L\"auchli,
\newblock {Light-Cone Dynamics After Quantum Quenches in Spin Chains},
\newblock Phys. Rev. Lett. {\bf 113}(18), 187203 (2014).

\bibitem{DSY}
B.~Doyon, H.~Spohn, and T.~Yoshimura,
\newblock A geometric viewpoint on generalized hydrodynamics,
\newblock Nucl. Phys.  {\bf B926}, 570--583 (2018).

\bibitem{ourroaming}
M.~Mazzoni, O.~Pomponio, O.~A. Castro-Alvaredo, and F.~Ravanini,
\newblock {The Staircase Model: Massless Flows and Hydrodynamics},
\newblock J. Phys. {\bf A54} 404005 (2021).

\bibitem{wiki} {https://en.wikipedia.org/wiki/Crown\_shyness}

\end{thebibliography}
\end{document}